\documentclass[a4paper,11pt]{article}
\usepackage{jheppub} 
\usepackage{lineno}
\usepackage{wrapfig}
\usepackage{caption}
\usepackage{subcaption}

\usepackage{microtype}

\usepackage{comment}

\newcommand{\GH}[4]{ {_2F_1\left(#1,#2;#3;#4 \right)} }
\newcommand{\Hl}[7]{ {Hl\left(#1,#2;#3,#4,#5,#6;#7 \right)} }
\newcommand{\GA}[1]{ {\Gamma \left( #1 \right) } }
\newcommand{\DG}[1]{ {\psi \left( #1 \right) } }

\newcommand{\dd}{\mathrm{d}}

\DeclareMathOperator{\pr}{\partial}

\title{\boldmath Light-like retarded correlators and the horizon}

\author{Justin R. David, Leonard Schwarze}
\affiliation{Centre for High Energy Physics, \\
Indian Institute of Science, \\
C. V. Raman Avenue, \\
 Bangalore 560012, India.}
\emailAdd{justin@iisc.ac.in, leonards@iisc.ac.in}

\abstract{
We show that  retarded correlators  in conformal field theories of 
  scalar primary operators 
   evaluated at finite temperature using $AdS/CFT$  scale anomalously at large light like momenta. 
   The anomalous scaling depends on the curvature  of the horizon. 
  Setting the momenta equal to the frequency $\omega$, the retarded correlator   for black holes with planar horizons 
  scales as $\omega^{\frac{2}{d+2} (2\Delta - d) }$ as opposed to the scaling behaviour of $\omega^{2\Delta - d}$  expected by dimensional analysis at generic fixed momenta and large frequencies. $\Delta$ is the dimension of the primary and $d$,  the number of space-time  dimensions.  
  For black holes with spherical and hyperbolic horizons when the frequency squared equals the Casimir along the horizon, the retarded correlator scales as $\omega^{\frac{2\Delta -d}{2} }$. 
  We establish this using exact results  in $d=2$,  and  a numerical analysis of the Heun equation for the  $AdS_5$
  planar black hole and finally  using the WKB approximation in general $d$.  The exact result for  black holes with hyperbolic horizon 
 as well as the  analysis at large $d$   provides additional checks for  the anomalous scaling behaviour. 
 Finally we evaluate  the anomalous scaling exponent for the stress tensor correlator in all its 
 3 channels. 
 }

\begin{document}
\maketitle
\flushbottom

\section{Introduction}
\label{sec:intro}

The study of retarded correlators  of various primary operators  in AdS/CFT  has  
given us several important insights for strongly coupled conformal field theories as well as 
quantum gravity.  The recipe to evaluate them in the AdS/CFT correspondence  was first introduced in 
\cite{Son:2002sd}.  Subsequently, the properties of these correlators at small frequencies and momenta 
and related developments 
allowed the exploration of the transport properties of strongly coupled CFT's at finite temperature 
via holography \cite{Kovtun:2003wp,Hubeny:2011hd}. 
 Indeed  the famous observation of the universality of the shear viscosity to 
entropy density in \cite{Kovtun:2004de}  came about by a study of  the  retarded correlator corresponding to the shear 
component of the stress tensor.  
There was then a concerted effort to study various strongly coupled phenomena in field theories using 
holography which was largely driven by the computation of retarded correlators  in several examples, see \cite{Hartnoll:2009sz} for a review. 
For most studies the low frequency and momenta properties of these correlators played a crucial role.  
Coupled with this development, there was a interest in understanding the analytical structure of the 
thermal correlators  obtained from holography \cite{Hartnoll:2005ju,Festuccia:2005pi}. 
These early 
 studies revealed that  holographic retarded correlators  are meromorphic  in the frequency plane and have poles 
 when the frequency coincides with the  quasi-normal modes of the black hole \cite{Horowitz:1999jd,Berti:2009kk}.

Motivated by developments in conformal bootstrap  there 
 has been renewed interest in the study of both  retarded correlators  and more general correlators 
at finite temperature \cite{Iliesiu:2018fao,Petkou:2018ynm,Gobeil:2018fzy,Alday:2020eua,Dodelson:2023vrw,David:2023uya,Bhattacharya:2025vyi,Buric:2025anb,Barrat:2025nvu,Buric:2025fye,Barrat:2025twb,Jia:2025jbi,Jia:2026ryl}
The recent  focus  is on the 
analytic structure of the thermal correlators and exact properties both in position space  in the complex frequency plane.
Most of the studies regarding the analytical structure of  holographic correlators  in the 
complex frequency plane keeps the momentum  fixed or 
zero momentum.

In this paper our main focus is on the behaviour of holographic retarded correlators at large frequencies in the light like limit. 
To begin let us define the correlator of interest, consider the  case in which the 
conformal theory in on the plane $R^{d-1}$ so that it has spatial translational invariance. 
\begin{eqnarray}
G^{R}( \omega, \vec k ) = -i \int d^{d} x e^{i \omega t - i \vec k \cdot \vec x } \theta ( t) \langle [\hat O(x) , \hat O(0 ) ] \rangle_\beta
\end{eqnarray}
Here ${\cal O }$ is a scalar conformal primary of dimensions $\Delta$, the  expectation value of the commutator is 
taken on the thermal state at inverse temperature $\beta$. 
The usual strategy to evaluate the retarded correlator in field theory is to Fourier transform the Euclidean 
two point function 
\begin{eqnarray}
G^E( \omega_E, \vec k ) = \int_0^\beta d\tau \int(dx)^{d-1}
 e^{- i \omega_E \tau -  i \vec k \cdot x} \Big\langle T_E \big(  \hat O( \tau, \vec x)  , \hat O (0) \big)   \Big\rangle_{\beta}  
\end{eqnarray}
where $T_E$ refers to time ordering in Euclidean time $\tau$, and $\omega_E$ are Matsubara frequencies. 
Then the retarded correlator is determined by the  analytical continuation of the Euclidean correaltor 
by  the replacement
$\omega_E = - i \omega$, 
 such that 
 \begin{eqnarray}
  \label{analyticalcont}
G^R( \omega, \vec k ) = - G^E( -i\omega , \vec k ) 
\end{eqnarray}
This ensures the condition that the retarded correlator evaluated at imaginary Matsubara frequencies 
reduces to its  Euclidean counterpart.
\begin{eqnarray}
G^R( \frac{2\pi i n }{\beta}, \vec k ) = - G^E( \frac{2\pi  n}{\beta}, \vec k ) 
\end{eqnarray}

At zero temperature, since the Euclidean two point function is known 
for any  conformal field theory, and is given by 
\begin{eqnarray}
\Big\langle T_E \big(  \hat O( \tau, \vec x , \hat O (0) \big)  \Big\rangle_{\beta =0 }   = 
\frac{1}{( \tau^2 + \vec x^2)^\Delta} 
\end{eqnarray}
where $\Delta$ is the conformal dimension of the operator $\hat O$ . 
Then evaluating the Fourier transform and analytically continuing,  we find that the retarded correlator is 
\begin{eqnarray}
G^{R}( \omega, \vec k )|_{\beta = 0} \sim  ( \omega ^2 - \vec k^2)^{ \frac{ 2 \Delta - d}{2} } 
\end{eqnarray}
This result can also be argued from using dimensional analysis and from the  Lorentz symmetry on $R^{1, d-1}$ 
at zero temperature.  We have also reviewed the derivation of this result from holography in 
appendix . 

At non zero temperature, 
we expect at least for generic momenta $\vec k$, but at large frequencies $\omega \gg T$  the short distance 
OPE to determine the dominant behaviour \cite{Caron-Huot:2009ypo}. 
The Euclidean correlator then can be approximated to by 
\begin{eqnarray} \label{ope1}
\Big\langle T_E   \big( \hat O( \tau, \vec x)  , \hat O (0) \big)  \Big\rangle_{\beta}    \sim  \frac{1}{( \tau^2 + \vec x^2)^\Delta} 
+ O\Big(  \frac{1}{( \tau^2 + \vec x^2)^{\Delta + \Delta_{O'} }  }  \Big) 
\end{eqnarray}
where $\Delta_{O'}$ refers to the lowest  conformal dimension of the operator which appears in the OPE. 
Therefore in this limit we expect the  retarded correlator to grow as  \footnote{When $\Delta$ is an integer 
$G^R( \omega, \vec k ) \sim  ( \omega ^2 - \vec k^2)^{ \frac{ 2 \Delta - d}{2} } \log ( \omega^2 - k^2 ) $}.
\begin{eqnarray} \label{leadterm}
\lim_{\omega\rightarrow \infty} G^{R}( \omega, \vec k )|_\beta  \sim (  \omega^2 - k^2 )^{ \frac{2\Delta - d}{2} }
\end{eqnarray}
As expected 
this behaviour  can be also be shown also from
evaluating the retarded Greens function holographically by solving the minimally coupled scalar  in a planar
$AdS_{d+1}$ black hole and taking
the large frequency limit  at generic momenta values of the momenta.  We will review this result subsequently. 

Inspite of the extensive literature, there has been very few studies of holographic retarded correlators
on the light cone or when $\omega = |k|$ with $\omega \rightarrow \infty$.  
Retarded correlators on the light cone are necessary when one evaluates 
on shell processes like photon and dilepton emission  rates or product of gravitational waves  at finite temperature
\cite{Caron-Huot:2006pee,Castells-Tiestos:2022qgu}. 
In fact, the only instance of anomalous scaling of the retarded correlator  that   we are aware of  
is in equation (3.19) 
of \cite{Caron-Huot:2006pee}
for the correlator of the $U(1)$ current in $AdS_5$. 

From  (\ref{leadterm}), we expect  the leading which originated from the first term in OPE expansion to vanish. 
Since at finite temperature, we do not have the Lorentz invariance, it is possible that sub-leading 
powers of frequency play a role in this light cone limit. 
In this paper we show that expectation is indeed true and we obtain the anomalous scaling 
of the retarded correlators on the light cone. 
There are two possible scaling behaviours that we find one for the black hole with a planar horizon and a different 
scaling law for a black hole with hyperbolic or spherical horizon. 

For the planar black hole we see that 
\begin{eqnarray} \label{lim1}
\lim_{\omega\rightarrow \infty} G^{R}( \omega, \vec k )|_{\beta, \omega = |\vec k | } = C
\omega^{(2 \Delta - d)\frac{2}{ d +2} }
\end{eqnarray}
In this and subsequent equations of the introduction frequency is measured in units of the termperature, that is 
$\omega \rightarrow \frac{\omega}{2\pi T}$.   The constant 
$C$ contains the necessary dependence on temperature to ensure 
the retarded Greens function has the correct scaling dimensions. 
We establish this, first by studying the $d=2$ case or the planar BTZ black hole in $AdS_3$ for which 
the retarded correlators for a conformal primary can be evaluated exactly both in the CFT and from holography, 
We then examine the planar black hole in  $d=4$  and study the holographic correlator numerically by means of 
the solutions to the Heun equation and establish the above scaling behaviour. 
Finally we use the WKB method to solve the wave equation and demonstrate the above scaling law for a planar 
black hole in $AdS_{d+1}$. 
The WKB method is general and shows that the scaling behaviour does not change for the Reissner-Nordström 
black hole with planar horizon.

Further using the WKB method we show that for black holes with spherical horizons as well as hyperbolic horizons
obey the following scaling behaviour when the frequency $\omega$ is large and obeys the dispersion relation 
\begin{eqnarray}
\omega^{* \, 2 }  = l ( l + d-2)         , \qquad\qquad             \hbox{spherical horizons}, \\ \nonumber
\omega^{*\, 2} = \lambda^2  + \big( \frac{ d-2}{2} \big)^2 , \qquad\qquad \hbox{ hyperbolic horizons} 
\end{eqnarray} 
Here $l$ large positive integer and $\lambda$ is a large real number. 
then 
\begin{eqnarray}
\lim_{\omega^* \rightarrow \infty} G^R( \omega^*,  l )|_{\beta}   =  C( \omega^*)^{\frac{2\Delta - d}{2}} , \\ \nonumber
\lim_{\omega^* \rightarrow \infty} G^R( \omega^*,  \lambda  )|_{\beta}  =  C ( \omega^*)^{\frac{2\Delta - d}{2}} 
\end{eqnarray}
as opposed to the usual scaling behaviour, $\omega^{2\Delta - d}$  at generic $l, \lambda$. 
Again, this scaling behaviour remains the same if black holes with such horizon topology also have charge. 

To demonstrate that such anomalous scaling behaviour is present in other retarded Green's function, we study the 
vector and sound mode of the planar black hole  in arbitrary dimensions using the  WKB approximation on the light cone
We find the following 
\begin{eqnarray} \label{lim2}
\lim_{\omega^\star\rightarrow\infty} G^{R}( \omega^\star, \vec k )_{\beta, \omega^\star = |k|} =  C\left(\omega^\star\right)^\frac{ 4d}{d+2}, \qquad \qquad \hbox{vector mode}, \\ \nonumber
\lim_{\omega^\star\rightarrow\infty} G^{R}( \omega^\star, \vec k )_{\beta, \omega^\star = |k|} = C   \left(\omega^\star\right)^\frac{ 6d}{d+2}. \qquad \qquad \hbox{sound mode}
\end{eqnarray} 

As a final consistency check, we note that in the strict light like limit, the equation of motion for the minimally coupled scalar, 
shear, vector, and the sound mode of the vector fluctuations
 is exactly solvable in the large $d$ limit. We use this exact solution to obtain the 
large frequency limit and show that it is consistent with the $d\rightarrow\infty$  limit of 
(\ref{lim1}), (\ref{lim2}),

 We now briefly compare the light cone limit that we have take in to other light cone limits in the literature. 
 In \cite{Hubeny:2006yu,Dodelson:2020lal,Dodelson:2023nnr} , 
the authors study the light cone singularities in position space of the retarded coorrelator and 
identify null geodesics which are responsible for this divergence. It would be interesting to see 
if  we can relate the sub-leading scaling behaviour on the light cone in momentum space  to a contribution of 
a null geodesic in position space. 
In another series of works  \cite{Huang:2022vet,Esper:2023jeq} 
in which a near light cone limit was studied for stress tensor correlators. 
In this limit, the authors showed that position space retarded correlators for the stress tensor are determined by 
3 universal coefficients. This near light cone limit \footnote{See equation (3.8) of \cite{Esper:2023jeq} }, 
also involves  a scaling 
of the radial co-ordinate which is not present in the strict light cone limit in momentum space discussed in this paper. 

The organisation of the paper is as follows. 
In section \ref{planar-bhsection}, we study the behaviour of the retarded correlator of the minimally coupled sclar 
 for light-like
momentum in the planar black hole. It includes the discussion of $d=2, d=4$ and  the derivation of 
the scaling behaviour in (\ref{lim1}) using the WKB approximation.  
In section \ref{sec-sphere}, we generalise the WKB method to obtain the retarded correlator with light-like momenta for 
 the case of black holes 
with spherical and hyperbolic horizons. 
We cross check the WKB answer with the exact result for  the retarded Green's function for  hyperbolic  black holes.
In section \ref{stress} we derive the anomalous scaling exponent  of the 
 retarded correlator of the stress tensor in its  3 channels in arbitrary dimensions
using the WKB approximation.
In section \ref{sec:large-d} we show that in the large $d$ limit, the  ODE's  to derive the retarded correlator
at light-like momenta  
for minimally coupled scalar and the stress tensor channels in the planar black hole can be solved in 
terms of hypergeometric functions.  Section \ref{conclusion} contains our conclusions and discussions. 
The appendix \ref{appena} reviews the result for the retarded correlator in pure $AdS$ to highlight the difference 
seen for the correlator with light-like momenta. 
In appendix \ref{borelresum} we use Borel resummation of the momentum space OPE to obtain the anomalous
scaling of the retarded correlator  in $d=2$ for an operator with conformal dimensions $(1, 1)$.

\section{Light-like limit  in the planar black hole}
\label{planar-bhsection}
In this section we  study the light-like limit for retarded correlators  for a scalar operator of dimension $\Delta$ 
evaluated in the planar  $AdS_{d+1}$ black hole. We choose  $\Delta$ such that it is not an integer. 
We begin with the setup in $d=2$ for the correlator can be evaluated exactly both from holography and CFT, and we demonstrate that in the light-like limit the correlator scales anomalously. In this  case it is proportional to 
$\omega^{\Delta - 1}$ as opposed to the generic  scaling  behaviour  $\omega^{2\Delta-2}$. 
One of the byproducts of our study is the evaluation of the Fourier transform of the Euclidean thermal two point function of an operator of dimension  arbitrary $\Delta$ using results developed in  string perturbation theory  in \cite{Kawai:1985xq}\footnote{This Fourier transform was  performed for  integer $\Delta = 1, 2 $ and it was  argued to agree with the holographic result  for arbitrary $\Delta$ in \cite{Son:2002sd}. }. 
 
We then move to $d=4$   for which the minimally coupled  massive scalar  obeys the Heun equation \cite{Starinets:2002br}.  We use numerical methods to obtain the connection formula of the Heun equation, which relates the in going solution 
at the horizon and solution at the boundary and obtain the retarded correlator. 
We then study the correlator at large light-like frequencies  numerically 
and show that the retarded correlator scales  as $\omega^{\frac{ 2\Delta - 4}{3}}$ which is sub-leading 
compared to the generic behaviour $\omega^{2\Delta - 4 }$. 

Finally we use  the WKB  method to solve the minimally coupled scalar equation and obtain the retarded correlator in the 
light-like limit for arbitrary $d$.
This reveals the presence of the anomalous scaling law $\omega^{\frac{ 2 ( 2\Delta - d) }{d+2 }}$.

Before we begin, we set up the preliminary background for our analysis. 
The action for the minimally coupled scalar in given  by 
\begin{eqnarray}
S = -\frac{1}{16\pi G_N} \int \dd^{d+1}x \sqrt{-g}  \Big[
\frac{1}{2} ( \nabla^\mu \phi \nabla_\mu \phi + M^2 \phi^2) \Big].
\end{eqnarray}
In asymptotically $AdS_{d+1}$ spaces, the minimally coupled scalar is dual to an operator ${\cal O}$ whose conformal dimension
is related  to the mass by 
\begin{eqnarray} \label{mass-dim-rel}
\Delta ( \Delta - d) =  R^2 M^2. 
\end{eqnarray}
For standard quantization of the scalar field in $AdS_{d+1}$, the conformal dimension $\Delta$
 is identified with the greater root of 
the above equation,  $\Delta = \Delta_+$. 
The background metric we consider is the planar Schwarzschild black hole in $AdS_{d+1}$
\begin{equation} \label{planarmetric}
  \dd s^2 = \frac{R^2}{z^2}\left(-f(z)\dd t^2 + \frac{\dd z^2}{f(z)} + \dd \vec x^2\right), \quad f(z) = 1 - \frac{z^d}{z_0^d}, \quad z_0 = \frac{\beta d}{4 \pi}. 
\end{equation}
Here $\beta$ is the inverse temperature of the black hole and $z$ is the radial co-ordinate, with $z=z_0$ the horizon and 
$z=0$ the boundary of $AdS$.  $R$ is the Radius of $AdS$. The equation of motion of the minimally coupled scalar is
\begin{equation} \label{mincoup}
  \frac{1}{\sqrt{-g}}\pr_\mu\left( \sqrt{-g} g^{\mu \nu} \pr_\nu \phi \right) - M^2 \phi = 0,\quad g = \det g_{\mu \nu}.
\end{equation}
The retarded Green's function is found by solving the equations of motion with in going boundary conditions 
at the horizon and examining the behaviour of the solution at the boundary. 
At the boundary, the solution behaves as 
\begin{eqnarray}
\phi(z, t, \vec x)  \approx  z^{\Delta_-}  \phi_0( t, \vec x) \big( 1+ O( z^2)  \big) + 
z^ {\Delta_+} \phi_1(t, \vec x) \big( 1+ O(z^2) \big), 
\end{eqnarray}
where $\Delta_{\pm}$ are the two roots of the equation (\ref{mass-dim-rel}). 
Due to the translational invariance of the metric (\ref{planarmetric}), 
the solution can be expanded in Fourier modes as follows
\begin{equation}
  \phi(z, t, \vec x) = \exp[i  2\pi T(\vec k  \cdot \vec x - \omega \cdot  t)] \phi(z, \omega, \vec k  ).
\end{equation}
Here $T = \frac{1}{\beta}$ is the temperature of the black hole. Note that $(\omega, \vec k ) $ are dimensionless frequency and momentum.
The retarded Greens function is then given by \cite{McGreevy:2009xe}
\begin{eqnarray} \label{formularc}
G^R(\omega, \vec k   ) =  ( 2 \Delta - d)  
 \frac{R^{d-1} }{16\pi G_N}  \frac{\phi_1( \omega,  \vec k ) }{\phi_0( \omega, \vec k ) }, 
\end{eqnarray}
where $\Delta = \Delta_+$.  Observe that the dimensionless pre-factor measures Newton's constant in units of the radius of $AdS$ and gives the measure of the 
degrees of freedom. For example, in the case of $AdS_5\times S^5$, the pre-factor is proportional to $N^2$, the rank of the 
gauge group of the dual theory using the relation between radius of $AdS$, the Newton's constant and the 
t'Hooft coupling \cite{Maldacena:1997re}. 
For
 $d=2$, we see this factor is proportional to the central charge of the CFT using the Brown-Henneaux formula \cite{Brown:1986nw}.
Since the scalar field is dimensionless, the ratio has the dimension of $\Delta_+ - \Delta_- = 2 \Delta - d$,  the 
right scaling dimension for the  Fourier transform of the retarded correlator  of an operator of conformal dimension 
$\Delta$.  
To solve and analyse the solution of the minimally coupled scalar in the planar $AdS_{d+1}$ black hole in (\ref{mincoup}), it is 
useful to introduce the co-ordinate
\begin{eqnarray}\label{dcoordinate}
w = \Big( \frac{z}{z_0}  \Big)^d. 
\end{eqnarray}
In terms of this coordinate  the Fourier modes obey the equation 
\begin{equation}
  \label{eq:EoM-d}
  \phi''(w)+\frac{1}{w-1}\phi'(w) + \left[ w^{2/d}\frac{(w-1)k^2+\omega^2}{4 (w-1)^2 w^2} + \frac{h(h-1)}{(w-1) w^2} \right] \phi(w) = 0, 
\end{equation}
where
\begin{eqnarray}
h = \frac{\Delta}{d}. 
\end{eqnarray}
Here we have suppressed the dependence of the $(\omega, \vec k ) $ in $\phi$.

We will now proceed to study the retarded correlator (\ref{formularc}) in the light-like limit for 
$d=2, 4$, and for arbitrary $d$ using the WKB approximation.

\subsection{The correlator for $d=2$}

In this section we analyse the retarded correlator of a minimally coupled scalar in the $AdS_{3}$ planar black hole in the light-like limit. As a model that is exactly solvable, it provides valuable insights on the analytic structure of the correlator, with an interpretation on both the CFT side as well as holography. We will first compute the correlator holographically using the exact solution in terms of hypergeometric functions. We then proceed to present a CFT side analysis by computing the Fourier transform of the known Euclidean correlator.

\subsection*{BTZ Correlator from Holography}

The evaluation of the retarded correlator for the BTZ black hole has been done earlier 
\cite{Son:2002sd,Iqbal:2009fd}, but we review it here 
for completeness. 
We consider the  minimally coupled scalar in the planar 
BTZ black hole, for this we examine the equation in (\ref{eq:EoM-d})  with   $d=2$.

The resulting equation of motion is
\begin{equation}
  \label{eq:d=2}
  \phi''(w) + \frac{1}{w-1} \phi'(w) + \left[\frac{(w-1)k^2 + \omega^2}{4(w-1)^2 w} + \frac{h(h-1)}{(w-1)w^2}\right] \phi(w) = 0.
\end{equation}
This equation admits a closed form solution in terms of hypergeometric function, the solution which obeys the in going 
boundary conditions at the horizon, $w=1$  is given by 
\begin{equation}
 \phi(w) = A^{(1)} w^h (1-w)^ {- i \frac{w}{2} }  \; 
  \GH{h - i \frac{\omega+k }{2}}{h - i \frac{\omega - k }{2}}{1-i \omega}{1-w}.
\end{equation}
Using the connection formula of the hypergeometric function 
 to relate this solution as an expansion at boundary $w = 0 $, we obtain 
\begin{eqnarray}
\phi ( w) &=&  A^{(1)}  \left[  \frac{ \Gamma( 2h - 1) \Gamma( 1- i \omega ) }{\GA{h - i \frac{\omega-k}{2}} \GA{h - i \frac{\omega+k}{2}}} w^{ 1- h } \Big( 1 +  O(w )   \Big)  \right. \\ \nonumber 
& & \left. + \frac{ \Gamma( 1 - 2h ) \Gamma ( 1- i \omega) }{\GA{1-h - i \frac{\omega-k}{2}} \GA{1-h - i \frac{\omega+k}{2}}} w^{ h } \Big( 1 +  O(w )   \Big)  \right]
\end{eqnarray}
Now using the definition of the retarded correlator from (\ref{formularc}) with $d=2$, we obtain 
\begin{eqnarray} \label{eq:BTZ-correlator-holographic}
\left. G^R( \omega, k )\right|_{d=2} = 2( \Delta - 1)  \frac{R ( 2\pi T)^{2\Delta - 2} }{ 16\pi G_N}  
\frac{\GA{1-2h}\GA{h - i \frac{\omega-k}{2}}\GA{h - i \frac{\omega+k}{2}}}{\GA{2h-1}\GA{1-h-i\frac{\omega-k}{2}}\GA{1-h-i\frac{\omega+k}{2}}}. \nonumber \\
\end{eqnarray}
To arrive at this, we have used the relation between $w$ and $z$ in (\ref{dcoordinate}) for $d=2$ and the definition of 
 $z_0$ in (\ref{planarmetric}).

\subsection*{Scaling Limits of the BTZ Correlator}
We now proceed to investigate several kinematic regimes of eq. \eqref{eq:BTZ-correlator-holographic}. First, we investigate the asymptotic of $G_R(\omega, \kappa)$ for large $\omega$ and fixed $k$. The result can be obtained by using 
\begin{equation} \label{scale1}
  G_R(\omega,\kappa) \sim  (4)^{1-2h}\frac{\GA{1-2h}}{\GA{2h-1}} (-i \omega)^{4h-2} (2\pi T)^{ 4h -2} . 
\end{equation}
This is exactly the large $\omega$ scaling at generic momenta 
 one would expect form the OPE approximation of the thermal two point function as in (\ref{ope1}) and (\ref{leadterm}). 
 The finite temperature effects are washed out and the two point function scales as $2 \Delta - d$ with $d=2$ and $\Delta = d\, h$. 

Next, let us assume $k= \delta \, \omega$ with $0<\delta<1$. It is easy to compute the $\omega \to \infty$ asymptotic, which is 
\begin{equation} \label{scale2}
  G^R(\omega,\delta \, \omega) \sim 4^{1-2h} (1-\delta^2)^{2h-1} \frac{\GA{1-2h}}{\GA{2h-1}} (-i \omega)^{4h-2} (2\pi T)^{ 4h -2} .
\end{equation}
Again, the large $\omega$ scaling is the expected result of the zero temperature CFT. In fact, this regime is simply the fixed $k$ correlator dressed by the factor $(1-\delta)^{2h-1}$.

Note that the above scaling comes from the fact that all four $\Gamma$ functions of $G_R(\omega, \kappa)$ contribute to the large $\omega$ limit. Thus, the asymptotic may change if we suppress the contribution of some of the $\Gamma$ functions. A natural way to achieve this is to restrict the correlator to the light-like limit, i.e. to impose $k=\pm\omega$. Then, two of the $\Gamma$ functions become independent of $\omega$ and at $\omega \to \infty$ we get
\begin{eqnarray}
  \lim_{\kappa \to \pm \omega}G^R(\omega, \kappa) &=&    2( \Delta - 1)  \frac{R ( 2\pi T)^{2\Delta - 2} }{ 16\pi G_N}  
  4^{1-2h} \frac{\GA{\frac{1}{2}-h} \GA{h- i \omega}}{\GA{-\frac{1}{2} + h} \GA{1-h-i \omega}},  \\ \nonumber
   && \approx (4)^{1-2h} \frac{\GA{\frac{1}{2}-h}}{\GA{h - \frac{1}{2}}} (-i \omega)^{2h-1} (2\pi T)^{ 4h -2} .
\end{eqnarray}
Reinstating the dependence on $\Delta$, we see that the correlator scales as $\Delta - 1$ in $d=2$ in accordance 
with the  general formula (\ref{lim1}). 
Notably, the asymptotic limit for large $\omega$ with $\omega = \pm k$,  does not result in the zero temperature
 scaling and therefore does not arise from the leading term in the OPE expansion.  

We will now perform a detailed comparison of the above results to the retarded 
 two point function of a CFT at zero temperature in arbitrary dimensions $d$. 
 First let us examine the Euclidean correlator in position space. Here, the symmetries are strict enough to limit the shape of the two point function to be 
\begin{equation}
  G^E(\tau_1,  \vec x;  \tau_2 , \vec y ) \Big|_{T = 0} = \frac{\mathcal C}{  ( ( \tau_1 - \tau_2 )^2 + | \vec x - \vec y|^2 )^{\Delta}  },
\end{equation}
where ${\cal C}$ is the normalization. 
Using the translation symmetry to set $\tau_2 = \vec y  = 0$ and performing the Fourier transform 
and analytically continuing to Minkowski  frequencies, the standard result is \footnote{See for example in  \cite{Manenti:2019wxs} }
\begin{equation}
  G^R(\Omega, \vec K) \Big|_{T = 0} 
  = 
   {\mathcal C} \frac{\pi^{\frac{d}{2} } \Gamma( \frac{d}{2} - \Delta) }{ 2^{ 2\Delta - d} \Gamma( \Delta) } ( \vec K^2 - \Omega^2 ) ^{\frac{ 2 \Delta - d}{2} }. 
\end{equation}
Here $( \Omega, \vec K ) $ are dimension-full  frequencies and momenta. 
It is easy to see that  for $d=2$, the 
behaviour of the zero temperature retarded Green's function for large $\Omega$ at fixed momentum 
$\vec K$, or when $ \vec K= \delta \Omega$ coincides with (\ref{scale1}) and (\ref{scale2}) respectively.  This is expected from the 
OPE argument that at frequencies $\Omega \gg 2\pi T$, the thermal correlator reduces to that at  zero temperature.

It is clear that for $\Delta>\frac{d}{2}$ the zero temperature two-point function vanishes  on the light cone, 
 $\vec K ^2 - \Omega^2 =0$. The fact that the holographic result (\ref{eq:BTZ-correlator-holographic}) does not vanish on the light cone clearly indicates that sub-leading terms 
 must contribute on the light cone.

\subsection*{Retarded correlator from 2d CFT at arbitrary $h$}
We now proceed to show that the holographic result of the light-like scaling at finite temperature can be equivalently obtained from the CFT itself. We do so by performing the Fourier transform of the known Euclidean two point function \cite{Son:2002sd},
\begin{equation}
    G_E(x, \tau) = \frac{C_{\mathcal O} \left(\pi T\right)^{2 (h+\bar h)}}{\sinh\left( \pi T w \right)^{2h}\sinh\left( \pi T \bar w \right)^{2 \bar h}}.
\end{equation}
Here we have $w=x+ i \tau$.  To show that the scaling in the light-like limit can be derived from the CFT itself, we aim to perform the Fourier transform of this result, namely
\begin{equation}
    \label{eq:Fourier-Transform-2D}
    \begin{split}
    G_E(k, \omega_n) &= \int_{-\infty}^{\infty} \dd x \int_0^{\frac{1}{T}} \dd \tau e^{i k x} e^{i \omega_n \tau} G_E(x,\tau) \\
    &= C_{\mathcal O} \left(\pi T\right)^{2( h+ \bar h) } \int_{-\infty}^{\infty} \dd x \int_0^{\frac{1}{T}} \dd \tau \frac{e^{i k x} e^{i \omega_n \tau}}{\sinh\left( \pi T w \right)^{2h}\sinh\left( \pi T \bar w \right)^{2 \bar h}}.
    \end{split}
\end{equation}
This Fourier transform is easier to compute if we first unfold the cylinder by a map
\begin{equation}
    \mathbb R \times S^1 \longrightarrow \mathbb C\backslash \{0\},\quad
    e^{2 \pi T w}  \longmapsto z,\quad
    e^{2 \pi T \bar w}  \longmapsto \bar z, 
\end{equation}
mapping the thermal cylinder to the punctured complex plane. The factor of two in the exponent is chosen so that $\tau\in[0,\frac{1}{T}]$ corresponds to the angular variable in the complex plane. The power of this map can be fully appreciated when considering the definition of the hyperbolic sine in terms of exponential functions:
\begin{equation}
    \begin{split}
        \sinh(\pi T w) = \frac{e^{\pi T w} - e^{- \pi T w}}{2} = \frac{z^{\frac{1}{2}}-z^{-\frac{1}{2}}}{2}, \quad\sinh(\pi T \bar w) = \frac{e^{\pi T \bar w} - e^{- \pi T \bar w}}{2} = \frac{\bar z^{\frac{1}{2}}-\bar z^{-\frac{1}{2}}}{2}.
    \end{split}
\end{equation}
The Fourier kernel transforms under this map as
\begin{equation}
    e^{i k x} = (z \bar z)^{\frac{i k}{4 \pi T}},\quad e^{i \omega_n \tau} = \left(\frac{z}{\bar z} \right)^{\frac{\omega_n}{4 \pi T}},
\end{equation}
and for the Jacobian determinant we obtain
\begin{equation}
    |\det J_{(x,\tau)}| = \frac{1}{8 \pi^2 T^2 z \bar z}.
\end{equation}
Therefore, the Fourier transform \eqref{eq:Fourier-Transform-2D} becomes
{\small \begin{equation}
    \begin{split}
    \label{eq:Fourier-2D-zzbar}
    G_E(k, \omega_n) &= \frac{C_{\mathcal O} ( 2 \pi T)^{2(h+\bar h) }  }{8  \pi^2 T^2 } \iint_{\mathbb C\backslash \{0\}} \dd z \dd \bar z\, z^{h-1 + \frac{i k}{4 \pi T} + \frac{\omega_n}{4 \pi T}}(z-1)^{-2h} \, \bar z^{\bar h-1+\frac{i k}{4 \pi T} - \frac{\omega_n}{4 \pi T}}(\bar z - 1)^{-2 \bar h}\\
    &= \frac{C_{\mathcal O} ( 2 \pi T)^{2(h+\bar h) } }{8  \pi^2 T^2 } 
    \iint_{\mathbb C\backslash \{0\}} \dd z \dd \bar z\, z^{h-1 + \frac{i k}{4 \pi T} + \frac{\omega_n}{4 \pi T}}(1-z)^{-2h} \, \bar z^{ \bar h-1+\frac{i k}{4 \pi T} - \frac{\omega_n}{4 \pi T}}(1-\bar z)^{-2 \bar h}.
    \end{split}
\end{equation} }
Integrals of this type have been studied extensively in the context of equivalences between open and closed string amplitudes. Particularly, there is a relation due to Kawai, Lewellen, and Tye \cite{Kawai:1985xq}, that maps such integrals in the complex plane to products of contour integrals over the real line. They derive the relation
\begin{equation}
    \label{eq:KLT-relation}
    \begin{split}
    \iint \dd^2 z\, &z^{k_1 k_2 + n_{12}}(1-z)^{k_2 k_3 + n_{23}} \bar z^{k_1 k_2 + \bar n_{12}} (1-\bar z)^{k_2 k_3 + \bar n_{23}}\\
    = &-2 \sin(\pi k_2 k_3) \int_0^{1}\dd \xi \, \xi^{k_1 k_2 + n_{12}}(1-\xi)^{k_2 k_3 + n_{23}} \\
    &\times e^{i \pi k_2 k_3}\int_1^{\infty}\dd \eta\, \eta^{k_1 k_2+\bar n_{12}} (1-\eta)^{k_2 k_3 + \bar \eta_{23}},
    \end{split}
\end{equation}
by splitting the real and imaginary parts in $z = x '+ i y'$ and then deforming the $y'$ contour around the branch points onto the other axis. Here, the $k_i$ are the momentum space variables of the string amplitude and $n_{ij}$ are integers that deform the tachyonic four point function. The factor of $e^{i \pi k_2 k_3}$ stems from $|1-\eta|^\alpha = (1-\eta)^\alpha e^{-i \pi \alpha}$ for $\eta>1$. If the integrand in eq. \eqref{eq:Fourier-2D-zzbar} has no singularity at $z=\bar z = 0$ and if we can bring the exponents into the form required by \ref{eq:KLT-relation}, we may extend the integration domain to the full complex plane and use the above result to transform the integral over $\mathbb C$ into two Euler-type integrals. 
To use the above relation, the exponents of $z$, $\bar z$ and $1-z$, $1-\bar z$ must differ by integers to ensure single valuedness of the integrand. We have 
\begin{equation}
  (h-1 + \frac{i k}{4 \pi T} + \frac{\omega_n}{4 \pi T})-(\bar h-1 + \frac{i k}{4 \pi T} - \frac{\omega_n}{4 \pi T}) = (h - \bar h) + \frac{\omega_n}{2 \pi T}.
\end{equation}
Note that the spin $h-\bar h = s \in \mathbb Z/2$. For integer spin the bosonic Matsubara frequencies are $\omega_n = 2 \pi T n$. If $s\in \mathbb Z/2 \setminus \mathbb Z$, the fermionic frequencies are $\omega_n = 2 \pi T n + \pi T$. For the other exponent difference we have 
\begin{equation}
  -2h + 2\bar h = -2(h-\bar h) = -2 s \in \mathbb Z.
\end{equation}
Therefore, the exponent differences are indeed integers and the use of the above relation is justified.
We therefore write eq. \eqref{eq:Fourier-2D-zzbar} as
\begin{equation}
    \begin{split}
    G_E(k, \omega_n) &= -\sin(\pi k_2 k_3) C_{\mathcal O}e^{i \pi k_2 k_3}
   ( 2 \pi T)^{2 (h + \bar h - 1) } 
     \underbrace{\int_0^1 \dd \xi\, \xi^{h-1+\frac{i k+\omega_n}{4 \pi T}} (1-\xi)^{-2h}}_{T_1}\\
    &\times \underbrace{\int_1^\infty \dd \eta\, \eta^{\bar h-1+\frac{i k - \omega_n}{4 \pi T}}(1-\eta)^{-2 \bar h}}_{T_2}.
    \end{split}
\end{equation}
The integral $T_1$ can be immediately identified with the Euler integral of the first kind:
\begin{equation}
    T_1 = \mathrm B\left(h+\frac{\omega_n + i k}{4 \pi T}, 1-2h\right) = \frac{\GA{1-2h}\GA{h+\frac{\omega_n + i k}{4 \pi T}}}{\GA{1-h+\frac{\omega_n + i k}{4 \pi T}}}.
\end{equation}
To be precise, this integral converges only for $\Re(h)<\frac{1}{2}$, but the Euler beta function is analytically continued through its definition in terms of $\Gamma$ functions. On the level of integrals, this can be done by integrating along the Pochhammer contour $\mathcal C_P$. Whenever either of the branch points at the ends of the integration contour vanishes, the integral over $\mathcal C_P$ vanishes by Cauchy's theorem, resulting in the equation for the beta function being an indeterminate form. Interesting cases when this happens are for example $\Delta$ being positive integers. It is precisely this case that is, on the level of $\Gamma$ functions, represented by one of the $\Gamma$ running into a singularity. These integer $\Delta$ cases may be handled by by a suitable regularisation procedure, removing the singularity by e.g. a minimal subtraction scheme.

The integral $T_2$ can be mapped to an Euler integral of the first kind by the map $\eta \to \zeta = \frac{1}{\eta}$. The differential transforms as $\dd \eta = -\frac{\dd \zeta}{\zeta^2}$. Accounting for $(\zeta-1)^\beta = e^{-i \pi \beta}(1-\zeta)^\beta$ for $0<\zeta<1$, the result for $T_2$ is then
\begin{equation}
    \begin{split}
    T_2 &= e^{2 \pi i h}\int_0^1\dd \zeta\, \zeta^{h-1+\frac{\omega_n-i k}{4 \pi T}}(1-\zeta)^{-2h} = (-1)^{-2h} \mathrm B\left( h+\frac{\omega_n - i k}{4 \pi T},1-2h \right)\\
    &= e^{2 \pi i \bar  h}\frac{\GA{1-2 \bar h}\GA{h+\frac{\omega_n - ik}{4 \pi T}}}{\GA{1-\bar h+\frac{\omega_n-i k}{4 \pi T}}}.
    \end{split}
\end{equation}
Putting everything together and simplifying the prefactors then yields the Fourier transform of the correlator
\begin{eqnarray}\label{eq:BTZ-Correlator-FT2}
G_E( k, \omega_n)  &=& C_{\mathcal O}   ( 2 \pi T)^{2 (h + \bar h - 1) }  e^{- i \pi  s} 
\frac{1}{2i} \Big( e^{ i \pi ( h + \bar h) }  - e^{ - i \pi 2s }  e^{ -  i \pi ( h + \bar h) }  \Big)   \nonumber \\
&& \times \frac{ \GA{ 1-2h}
 \GA{h+\frac{\omega_n + i k}{4 \pi T}}  \GA{ 1- 2 \bar h} \GA{\bar h+\frac{\omega_n - ik}{4 \pi T}}}{\GA{1-h+\frac{\omega_n + i k}{4 \pi T}} \GA{1-\bar h+\frac{\omega_n-i k}{4 \pi T}}}. 
\end{eqnarray}
Observe that the phases in the first line have different behaviour for bosonic  fields  for which $s$ in an integer 
or fermionic fields for which $s$ is a half integer.  To arrive at this, we used the relations between the exponents in our integral and the ones defined in the KLT relation.
When the spin is zero,  $h=\bar h$, we obtain 
\begin{equation}
    \label{eq:BTZ-Correlator-FT}
    \begin{split}
    G_E(k,\omega_n) &=  C_{\mathcal O}   ( 2 \pi T)^{ 4h  - 2) } 
    \frac{\GA{1-2h}}{\GA{2h}} \frac{\GA{h+\frac{\omega_n + i k}{4 \pi T}} \GA{h+\frac{\omega_n - ik}{4 \pi T}}}{\GA{1-h+\frac{\omega_n + i k}{4 \pi T}} \GA{1-h+\frac{\omega_n-i k}{4 \pi T}}}.  
    \end{split}
\end{equation} 
As discussed before, this correlator is defined whenever $\Delta=2h \notin \mathbb N$. Otherwise, $\GA{1-2h}$ runs into a pole (the branch cut at the upper end of the integration domain vanishes) and one must regularise the expression. Subtracting only the divergent part, we find for $h=\frac{1}{2}$
\begin{equation}
    \label{eq:Euclidean-Correlator-h=1/2}
    G_E^{h=\bar h=1/2}(k,\omega_n) = -C_\mathcal O  \left[ 2 \gamma_E + \DG{\frac{1+n}{2}+\frac{i k}{4 \pi T}}+\DG{\frac{1+n}{2}-\frac{i k}{4 \pi T}} \right],
\end{equation}
and for $h=1$
\begin{equation}
\label{eq:Euclidean-Correlator-h=1}
  \begin{split}
    G_E^{h=\bar h=1}(k,\omega_n) = C_\mathcal O ( 2\pi T)^2 \bigg[ &(k^2+\omega_n^2) \left( 2\gamma_E + \DG{1+\frac{n}{2}+\frac{i k}{4 \pi T}}+\DG{1+\frac{n}{2}-\frac{i k}{4 \pi T}} \right)\\
    & -2 (k^2+\omega_n^2) + 4 \pi T \omega_n \bigg].
  \end{split}
\end{equation}
The $\gamma_E$ terms may be subtracted by modifying the subtraction scheme. Doing so makes eq. \eqref{eq:Euclidean-Correlator-h=1/2} coincide with the result in \cite{Son:2002sd}, up to an overall normalisation. The term $-2(k^2+\omega_n^2)$ is a local contact term. After removing this in eq. \eqref{eq:Euclidean-Correlator-h=1} the term $4 \pi T \omega_n$ appears in addition to the usually cited result \cite{Son:2002sd}. Importantly, eq. \eqref{eq:BTZ-Correlator-FT} is directly proportional to the correlator in eq. \eqref{eq:BTZ-correlator-holographic} after the analytical continuation $i \omega_n \to \omega$.
Thus, we have shown that the scaling in the light-like limit is present in a pure CFT analysis, at least in $d=2$. We stress that, although the term proportional to $\omega$ seems to be frequently dropped for $\Delta=2$, this term really emerges as a limit of a more general non-integer power, and should therefore have a non-trivial interpretation than just a contact term.

Before we leave this section, let us examine the general result (\ref{eq:BTZ-Correlator-FT2}) for arbitrary spins. 
It coincides with the result in \cite{Iqbal:2009fd}  obtained for retarded correlators of Dirac fermions in the BTZ black hole
 after 
the analytical continuation of $i \omega_n \rightarrow \omega$. 
More interestingly, 
note that in the light-like limit the  behaviour at large $\omega $ on the left moving light cone  $\omega = k$ is distinct 
form that of the right moving light cone $\omega = -k$. 
We obtain the following asymptotic behaviour of the retarded correlator
\begin{eqnarray}
\lim_{\omega \rightarrow \infty}G_R( k , \omega) \approx (- i \omega)^{2\bar h -1} ,\qquad\qquad k = \omega, ,   \\ \nonumber
\lim_{\omega \rightarrow \infty} G_R( k , \omega) \approx  (- i \omega)^{2h -1} ,\qquad\qquad k = -\omega, 
\end{eqnarray}
It will be interesting to show this behaviour in the bulk for arbitrary spins.

\subsection{Light-like limit in the $AdS_5$ planar black hole}
Until now, we have restricted our analysis to two-dimensional CFTs. Although in higher dimensions a CFT analysis is much more difficult, the case of $d=4$ is special in the sense that we can at least use semi-analytical methods. This is because the equation of motion of the minimally coupled scalar in the $AdS_5$ planar black hole is the Heun equation (see for example \cite{Starinets:2002br}). One semi-analytical approach is described in \cite{Dodelson:2022yvn}, where the authors use the analytical connection formulae developed in \cite{Bonelli:2022ten}, and approximate the Nekrasov-Shatashvili partition. For simplicity, we will use a different approach. If the Heun functions are to be evaluated well within the radius of convergence of their series expansion around a singular point, they can be computed to high precision reasonably fast. Thus, we will solve such problems by setting up the $2\times 2$ system of the connection problem arising from evaluation of the functions at two distinct points in the bulk, which can then be efficiently inverted. The equation of motion of the minimally coupled scalar in $\mathrm{AdS_5}$ is given by setting $d=4$ in eq. \eqref{eq:EoM-d}. Using the variable $u = \sqrt{w}$, one finds
\begin{equation}
  \label{eq:ODE-d=4}
  \phi''(u) + \frac{1+u^2}{u^3-u}\phi'(u) + \left[ \frac{(u^2-1)\kappa^2+\omega^2}{u(u^2-1)^2} + \frac{4h(h-1)}{u^2(u^2-1)} \right] \phi(u) = 0.
\end{equation}
This equation has singular points at $u=0$, $u=1$, $u=\infty$ and $u= -1 =: a$. Its $P-$symbol is 
\begin{equation}
  P\begin{Bmatrix}
    0 & 1 & a & \infty &\\
    2h & - \frac{i \omega}{2} & -\frac{\omega}{2} & 1 & \,;u\\
    2(1-h) & \frac{i \omega}{2} & \frac{\omega}{2} & -1 &
  \end{Bmatrix},
\end{equation}
which satisfies the Fuchsian condition. The $P-$symbol collects the local exponents at each of the singular points in each column. It can be identified with solutions to the Fuchsian equation when supplied with the accessory parameters. The standard form of the solution as reported in e.g. \cite{Maier_2006} is reached when the finite singular points have one vanishing exponent. Therefore, we can identify the canonical solution in terms of Heun functions after shifting the exponents in the $P-$ symbol as follows
\begin{equation}
  P\begin{Bmatrix}
    0 & 1 & a & \infty &\\
    0 & 0 & 0 & 2h - \frac{1+i}{2}\omega & \,;u\\
    2(1-2h) & i \omega & \omega & 2h - \frac{1+i}{2}\omega &
  \end{Bmatrix} = u^{-2h} (1-u)^{\frac{i \omega}{2}} (a-u)^{\frac{\omega}{2}} P\begin{Bmatrix}
    0 & 1 & a & \infty &\\
    2h & - \frac{i \omega}{2} & -\frac{\omega}{2} & 1 & \,;u\\
    2(1-h) & \frac{i \omega}{2} & \frac{\omega}{2} & -1 &
  \end{Bmatrix}.
\end{equation}
The canonical solutions in terms of $\psi(u) = u^{-2h} (1-u)^{\frac{i \omega}{2}} (1+u)^{\frac{\omega}{2}} \phi(u)$ can then be read off e.g. table 2 in \cite{Maier_2006}. The local solutions around the boundary are (rows $[0_+][1_+][a_+][\infty_+]$ and $[0_-][1_+][a_+][\infty_-]$ of table 2 in \cite{Maier_2006})
\begin{equation}
  \begin{split}
  \psi^{(0)}(u) = &A^{(0)}_{0}\,\Hl{a}{q}{\alpha}{\beta}{\gamma}{\delta}{u} + A^{(0)}_{1-\gamma}\, x^{1-\gamma} \\
   &\times \Hl{a}{q-(\gamma-1)(\delta a + \alpha+\beta-\gamma-\delta+1)}{\beta-\gamma+1}{\alpha-\gamma+1}{2-\gamma}{\delta}{u}.
  \end{split}
\end{equation}
Note that we have factored out the behavior at the singular points such that the regular solution corresponds to the normalisable mode and the one with local exponent $1-\gamma$ to the non-normalisable mode.
The local basis near the horizon is (rows $[1_+ 0_+][a_+][\infty_+]$ and $[1_- 0_+][a_+][\infty_-]$ of table 2 in \cite{Maier_2006})
\begin{equation}
  \begin{split}
    &\psi^{(1)} = A^{(1)}_{0} \Hl{1-a}{-q+\beta \alpha}{\alpha}{\beta}{\delta}{\gamma}{1-u} + A^{(1)}_{1-\delta} (u-1)^{1-\delta}\\
    &\times\Hl{1-a}{-q+(\delta-1)\gamma a + (\beta-\delta+1)(\alpha-\delta+1)}{\beta-\delta+1}{\alpha-\delta+1}{2-\delta}{\gamma}{1-u}.
  \end{split}
\end{equation}
The parameters of the solution are 
\begin{align}
  \begin{alignedat}{2}
  a &= -1, \quad &&q = \omega^2-\kappa^2 - \frac{1-i}{2}(4h-1)\omega\\
  \alpha &= 2 h - \frac{1+i}{2}\omega, \quad &&\beta =2 h - \frac{1+i}{2}\omega \equiv \alpha\\
  \gamma &= 4h-1, \quad &&\delta= 1-i \omega,
  \end{alignedat}
\end{align}
with the standard identity $\alpha+\beta+1=\gamma+\delta+\epsilon$.

\subsection*{Numerical Study}
\begin{figure}[t] 
  \begin{center}
  \includegraphics[width=.8\linewidth]{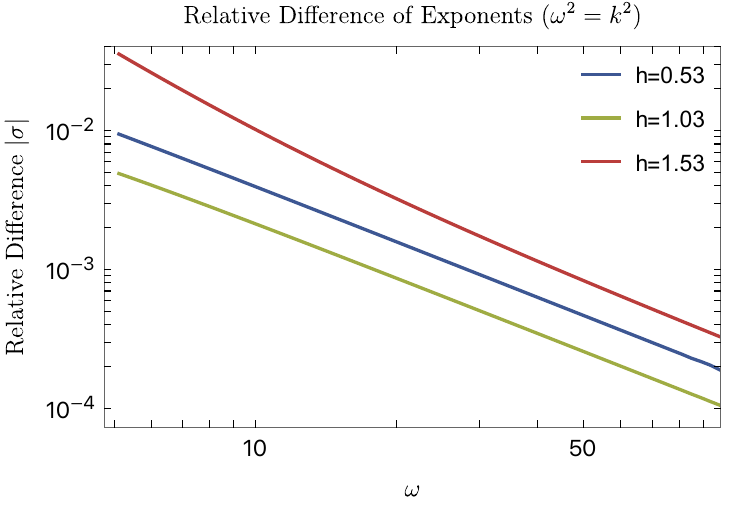}
  \end{center}
  \caption{The figure shows the relative difference of the large $\omega$ scaling exponent $\alpha$ to expected value of $\alpha_{\mathrm{exp}}=\frac{4}{3}(2h-1)$. The relative difference is defined by $\sigma = \frac{\alpha_{\mathrm{exp}}-\alpha}{\alpha_{\mathrm{exp}}}$. The exponent $\alpha$ is extracted via the logarithmic derivative $\alpha = \frac{\dd \log(|G(\kappa, \omega)|)}{\dd \log \omega}$.}
   \label{fig:lineplots-heun}
\end{figure}
We now make the numerical study more precise. The relevant singular points of eq. \eqref{eq:ODE-d=4} that we need to consider is the boundary ($u=0$) and the horizon ($u=1$). We already know that the solutions are Heun functions. The local expansions around the singular points only converge within their respective radii of convergence. Due to the singularity structure, the radius of convergence around the horizon and boundary in both cases equals one. We can immediately discard the irregular basis element around $u=1$, because we have defined the the regular solution to be infalling. Therefore, the connection problem we aim to solve is
\begin{equation}
     \begin{pmatrix}\psi^{(1)}_{\mathrm{regular}}\\0\end{pmatrix} = \begin{pmatrix}
    A & B\\
    0 & 0
  \end{pmatrix} \begin{pmatrix}
    \psi^{(0)}_\mathrm{regular}\\
    \psi^{(0)}_\mathrm{irregular}
  \end{pmatrix},
\end{equation}
i.e. we are only interested in half of the entries of the connection matrix. The strategy now is to evaluate this expression at two distinct points in the overlap region of validity of the horizon and the boundary $\mathcal D_{(0,1)} = D(0,1) \cap  D(1,1)$, where $ D(z,r)$ is an open disk of radius $r$ centred around $z\in \mathbb C$. The resulting linear system for the connection coefficients is then 
\begin{equation}
  \begin{pmatrix}
    \psi^{(1)}_{\mathrm{regular}}(z_1)\\
    \psi^{(1)}_{\mathrm{regular}}(z_2)
  \end{pmatrix} = \begin{pmatrix}
    \psi^{(0)}_{\mathrm{regular}}(z_1) & \psi^{(0)}_{\mathrm{irregular}}(z_1)\\
    \psi^{(0)}_{\mathrm{regular}}(z_2) & \psi^{(0)}_{\mathrm{irregular}}(z_2)
  \end{pmatrix} \begin{pmatrix}
    A\\B
  \end{pmatrix}, \quad (z_1=1/3,z_2=2/3) \in \mathcal D_{(0,1)}.
\end{equation}
If $1-\gamma = 4h-2$ is not an integer, there is no logarithmic branching at $u=0$, and no renormalisation must be performed. In these cases, the retarded two point function is directly proportional to the ratio $\frac{A}{B}$. Although this method is very simple to implement, it suffers from stability issues at large numerical values of the Heun parameters. We therefore choose to not exceed $\omega = 100$. The numerical results for $\kappa=\omega$ are shown in figure \ref{fig:lineplots-heun}.
We show the relative difference of the large $\omega$ exponent $\alpha$ to the exponent $\alpha_\mathrm{exp}=\frac{4}{3}(2h-1)$. We compute the exponent numerically using the logarithmic derivative $\alpha = \frac{\dd \log(|G(\kappa, \omega)|)}{\dd \log \omega}$, i.e. the plotted objective is $\sigma = \frac{\alpha_{\mathrm{exp}} - \alpha}{\alpha_{\mathrm{exp}}}$. We show the objective from $\omega=5$ to cut off any low frequency behavior as we wish to focus on the convergence at large $\omega$. The relative error between the numerical result and the proposed target exponent decays as a power law for large $\omega$. We therefore have used the analytical solution of the minimally coupled scalar in the $AdS_5$ planar black hole to show using a semianalytic method that the particular limit $\kappa = \omega$ does not yield the scaling for which $\alpha = 2 \Delta - 4$ but rather a modification thereof, with an exponent of $\alpha_{\mathrm{exp}} = \frac{4}{3}(2h-1) =\frac{1}{3} ( 2\Delta - 4) $. We shall explain the origin of this particular exponent through a WKB analysis in the bulk in the next section.

\subsection{WKB Analysis for the planar black hole in $AdS_{d+1}$}
\label{sec:planarbh}

As a starting point we shall consider the metric of a black hole in $AdS_{d+1}$
\begin{equation}
  \dd s^2 = \frac{R^2}{z^2}\left(-f(z)\dd t^2 + \frac{\dd z^2}{f(z)} + \dd \vec x^2\right).
\end{equation}
We consider blackening factors for planar geometries. We assume that in the domain of relevance there is a unique $z^\star$ such that $f(z^\star)=0$. In geometries with multiple horizons, $z^\star$ will consequently the outermost horizon. The geometries are asymptically $AdS$, i.e, $f(0)=1$, and the thermal nature is expressed via $f'(z^\star)\neq 0$.
Examples of the blackening factor are 
\begin{eqnarray}
 &&f(z) = 1 - \frac{z^d}{(z^\star)^d}, \phantom{+ q^2 z^{2d -2}\,\;} \qquad \hbox{ planar Schwarzschild black hole},  \\ \nonumber
 &&f(z) = 1 - \frac{z^d}{(z^\star)^d} + q^2 z^{2d -2}, \qquad \hbox{ planar Reissner-Nordström black hole}. 
\end{eqnarray}
Note that for the planar Reissner-Nordström black hole, the correction to the blackening factor due to the charge $q$ 
 are suppressed at the boundary $z=0$ compared to the   dominant the  Schwarzschild term $z^d$. 
Similar blackening factors occur for black hole solutions in higher derivative gravity where the perturbative corrections to the blackening factor due to the higher derivative terms are suppressed at the boundary 
\cite{Cai:2001dz,Myers:2010ru}. 
Since these solutions are asymptotically $AdS$, we have 
$f(0)=1$. 
The condition of the temperature is then 
\begin{equation}
  |f'(z^\star) |=  \frac{4 \pi}{\beta}.
\end{equation}
Consider the  Fourier component of equation of motion of a massive minimally coupled scalar by substituting 
\begin{equation}
\phi ( t, \vec x, z) = \phi ( z) e^{ i (   \vec K  \cdot x - i \Omega t ) }.  
\end{equation}
Then the  equation of motion of a massive minimally coupled scalar $\phi(z)$ reduces to 
\begin{equation}
  \phi''(z)+\left( \frac{1-d}{z}+\frac{f'(z)}{f(z)} \right) \phi'(z) + \left(\frac{\Omega^2 - K^2 f(z)}{f(z)^2} - \frac{d^2\,h(h-1)}{z^2 f(z)} \right) \phi(z) = 0.
\end{equation}
One may use the scaled dimensionless variable $w=\left(\frac{z}{z^\star}\right)^d$ together with the re-defintion $\phi(w) = f(w)^{-1/2} \psi(w)$ to obtain the following equation in Schrödinger form
\begin{equation} \label{scrodform}
  \psi''(w) + \left( w^{\frac{2}{d}-2} (z^\star)^2\frac{\Omega^2-K^2f(w)}{d^2f(w)^2} + \frac{f'(w)^2-2f(w)f''(w)}{4 f(w)^2} - \frac{h(h-1)}{w^2 f(w)}\right)\psi(w)=0.
\end{equation}
Now, let $|\vec K | = \delta \Omega$ for some fixed $\delta \in [0,1]$. Then, for large $\Omega$, the equation of motion becomes
\begin{equation}
  \psi''(w) + w^{\frac{2}{d}-2} (z^\star)^2 \frac{1-\delta^2 f(w)}{d^2f(w)^2}\Omega^2 \psi(w) = 0.
\end{equation}
We can now use the chain rule to express $z_0$ in terms of the temperature of the black hole via 
\begin{equation} \label{largefreqeq1}
  \frac{\dd f}{\dd z}\Big|_{z\to z^\star} = \frac{\dd w}{\dd z} \frac{\dd f}{\dd w} \Big|_{w \to 1} \implies z^\star = \frac{\beta d}{4 \pi} |f'(1)|.
\end{equation}
Introducing the dimensionless quantities $\omega = \frac{\Omega}{2 \pi T}$ and $\vec k = \frac{\vec K}{2 \pi T}$ then yields \footnote{We have verified that the term involving the second derivative $f''(w)$ in (\ref{scrodform}) 
is sub-dominant  for the charged planar black hole for $d>4$  in the domain (\ref{largefreqeq1}). }
\begin{equation} \label{largefreqeq}
  \psi''(w) + \underbrace{w^{\frac{2}{d}-2} \frac{(1-\delta^2 f(w))f'(1)^2}{4 f(w)^2}}_{Q(w)} \omega^2 \psi(w) = 0, 
  \qquad\qquad  \omega \gg1. 
\end{equation}

Our goal is to argue that the $\omega$ dependence of the correlator in the large $\omega$ limit is purely determined by the near boundary solution to the ODE. The basic argument is that the WKB solution transports $\omega$ merely as a phase and not in the pre-factors.  From the Schrödinger form  (\ref{scrodform}) we see that  the approximate equation in 
(\ref{largefreqeq}) is valid in the domain 
\begin{eqnarray} \label{regime0}
\frac{1}{\omega} \ll w  \ll1.
\end{eqnarray}
In this domain the WKB solution which  smoothly interpolates with the in-falling solution at the horizon is given by
\begin{eqnarray} \label{wkb}
\phi_{\mathrm{WKB}}(w)    =    \frac{A}{ Q(w)^{\frac{1}{4} } f(w)^{\frac{1}{2} } }   \exp\Big(  
 i \omega \int^w d t \sqrt{Q (t)  } \Big) .
\end{eqnarray}
Note that we are assuming that there are no turning points in the WKB potential  $Q(w)$. 
The frequency dependence of the WKB solution is only in the exponential. 
Let us first  verify that this  WKB solution matches the in falling solution at the horizon. 
At leading order near the horizon we have
\begin{equation}
  Q(w) = \frac{1}{4(1-w)^2} + \mathcal O((w-1)^{-1}).
\end{equation}
Substituting this expression for  $Q$  in (\ref{wkb}), we obtain 
that the WKB solution near the horizon is proportional to 
\begin{equation} \label{wkb0}
  \lim_{w \rightarrow 1 } \phi_{\mathrm{WKB}}(w) = A^{(1)} \, e^{-i \frac{\omega}{2} \log (1-w) }, 
\end{equation}
which  indeed matches the in-going solution at the horizon.  We have absorbed constants into  $A^{(1)} $. 
It is important to note that $A^{(1)} $ is independent of frequency. 

To obtain the  WKB solution near the boundary, we take the limit $w\rightarrow 0$. 
This  yields two separate cases for $\delta$ due to the condition $f(0)=1$.  
To see this, one may  expand $Q(w)$ for small $w$. 
\begin{equation} \label{Qexp}
  \begin{split}
  Q(w) &= w^{\frac{2}{d}-2}\left( \frac{(1-\delta^2f(0))f'(1)^2}{4 f(0)^2} - w \frac{(2-\delta^2 f(0))f'(0)f'(1)^2}{4 f(0)^3} \right) + \mathcal O(w^{2/d})\\
  &= w^{\frac{2}{d}-2} \left( \frac{(1-\delta^2)f'(1)^2}{4} - w \frac{(2-\delta^2)f'(0)f'(1)^2}{4} \right) + \mathcal O(w^{2/d}).
  \end{split}
\end{equation}
It is clear that the first term vanishes when $\delta=1$ but dominates for generic $\delta$. Thus, it is useful to treat the two cases separately.

\paragraph{Generic case:}
In the case $\delta \neq 1$, the leading WKB exponent is \footnote{Globally, the uncharged case can be evaluated as $$\int d t \sqrt{Q(t)} = \frac{d}{2} \vert f'(1) \vert \sqrt{1-\delta^2} w^{1/d} F_1\left( \frac{1}{d},-\frac{1}{2},1;1+\frac{1}{d};\frac{\delta^2}{\delta^2-1} w, w \right)$$ for $0<\delta<1$ with $F_1(a,b_1,b_2;c;x,y)$ being the Appell series $F_1$.}
\begin{equation} \label{integratalq}
  \int^w \dd t \sqrt{Q(t)} = \frac{d}{2}w^{\frac{1}{d}}\sqrt{1-\delta^2}f'(1).
\end{equation}
With $\sqrt{f(w)} = 1 + \mathcal O(w)$, we obtain
\begin{equation} \label{wkb1}
  \lim_{w \rightarrow 0} \phi_\mathrm{WKB,  \delta < 1}(w) = A^{(0)} w^{\frac{d-1}{2d}} e^{\frac{i \omega d}{2} w^{1/d}\sqrt{1-\delta^2}f'(1)} . 
\end{equation}
Here, we have absorbed the constant $\sqrt 2 ((1-\delta^2)f'(1)^2)^{-1/4}$ coming from the WKB prefactor into 
$A^{(0)}$.  The constant can have a dependence on $\omega$ through its phase. 
This is because we can choose to integrate the WKB exponent  from any point in the regime (\ref{regime0}), 
this leads to an integration constant in (\ref{integratalq}) which depends on the frequency linearly. 
Let us now match the WKB solution in (\ref{wkb}) to the solution of the differential equation (\ref{scrodform}) near the boundary. 
For this we can approximate the equation as 
\begin{equation}
  \psi''(w) + \left( \frac{\omega^2}{4}w^{\frac{2}{d}-2} (1-\delta^2)f'(1)^2 - \frac{h(h-1)}{w^2} \right)\psi(w) = 0,
\end{equation}
where we have retained the last term in \ref{scrodform})  which is dominant at the boundary. 
The solutions of the near boundary differential equation
are Bessel functions. For $\phi(w)$, we get
\begin{eqnarray} \label{bessel1}
 & &  \phi^{(0)}_{\delta<1}(w) = \sqrt{w}  \left( B J_{\nu}(\zeta) + C Y_{\nu}(\zeta) \right), \\ \nonumber
 & &  \nu = d \, h - \frac{d}{2} , \qquad  \zeta = \frac{\omega d}{2} w^{1/d}\sqrt{1-\delta^2}|f'(1)|. 
\end{eqnarray} 
For a fixed $w>0$ but large $\omega$ , we see that we need to  examine  the Bessel function at large 
argument. This solution then becomes 
\begin{eqnarray}
\lim_{\zeta\rightarrow \infty}   \phi^{(0)}_{\delta<1}(w)  
 & & = B w^{\frac{d-1}{2d} } \cos \Big(  \frac{\omega d}{2} w^{1/d}\sqrt{1-\delta^2}f'(1)  - \frac{\pi}{4} ( 2\nu + 1)  \Big)  \\ \nonumber
& & \qquad  +   C w^{\frac{d-1}{2d} } \sin \Big(  \frac{\omega d}{2} w^{1/d}\sqrt{1-\delta^2}f'(1)  - \frac{\pi}{4} ( 2\nu + 1)  \Big) 
\end{eqnarray}
where we have absorbed additional constants in $B, C$.
It is easy to see comparing (\ref{wkb1}), we  can solve for $B, C$ in terms of $A^{(0)}$. 
Furthermore it is important to note that the frequency dependence in $A^{(0)}$ is at most through a phase and consequently  no frequency dependence in $B, C$ depends on frequency through a common phase due to the 
WKB transport. 
Now that we have determined $B, C$, we can evaluate the retarded correlator. 
For this we evaluate the Bessel functions around $w=0$. 
This results in 
\begin{eqnarray}
\lim_{w\rightarrow 0 }   \phi^{(0)}_{\delta<1}(w)   = B^{(0)} \omega^{ d \, h } w^h  + C^{(0)} \omega^{d - d h} w^{ 1- h}. 
\end{eqnarray}
Here we have absorbed frequency independent constants in $B^{(0)}$ and $C^{(0)}$. 
To find the  retarded Green's function we need to evaluate the ratio of the normalisable mode to the 
non-normalisable mode. Taking this ratio cancels the frequency dependence of the phase of the constants
 $B^{(0)}$ and $C^{(0)}$ resulting in 
\begin{eqnarray}
\lim_{\omega \rightarrow \infty} 
G^R ( \omega,  |\vec k| = \delta \omega ) \approx \frac{\omega^{dh}}{ \omega^{ d-dh} } = \omega^{2\Delta - d}. 
\end{eqnarray}
This is the expected scaling behaviour for large $\omega$ and generic momenta.  This analysis also covers the 
case of fixed $\vec k$ for which $\delta = 0$.

\paragraph{Light-like case:}
One can now repeat the near-boundary analysis for the case $\delta=1$ \footnote{The global WKB exponent at $\delta=1$ and for the uncharged black hole is $\int \sqrt{Q(t)}d t = \frac{\vert f'(1) \vert}{2} B_w\left(\frac{1}{2} + \frac{1}{d}, 0 \right)$ with the incomplete Euler beta function $B_z(a,b)$.}. 
The near horizon behaviour is identical  to that found  for the generic $\delta$ and is given by (\ref{wkb0}). 
However the near boundary analysis differs. 
Now it is the sub-leading term in the expansion of $Q(w)$ given in (\ref{Qexp}) which contribute as $1-\delta^2 = 0$. 
We have 
\begin{eqnarray}
\lim_{w\rightarrow 0 } Q(w)_{\delta = 1}  = - \frac{1}{4} w^{\frac{2}{d} - 1}  f'(0) f'(1)^2  + O( w^{\frac{2}{d}} ) . 
\end{eqnarray}
 In this case, the WKB solution near the boundary  takes the form
\begin{equation}
 \lim_{w\rightarrow 0 } 
  \phi^{(0)}_{\mathrm{WKB}, \delta=1}(w) = w^{\frac{d-2}{4d}} \big( A^{(0)\prime} e^{\frac{i \omega d}{d+2} w^{\frac{1}{d} + \frac{1}{2}}\sqrt{-f'(0)f'(1)^2}}  \big). 
\end{equation}
The differential equation near the boundary becomes 
\begin{eqnarray} 
  \psi''(w) - \left( \frac{\omega^2}{4}w^{\frac{2}{d}-1} f'(0)f'(1)^2 + \frac{h(h-1)}{w^2} \right)\psi(w) = 0,
\end{eqnarray}
The near-boundary solution is 
\begin{eqnarray} \label{bessel2}
& & \phi^{(0)}_{\delta=1}(w) = \sqrt{w}  \left( B' J_{\nu'}(z) + C' Y_{\nu'} (z) \right),  \\ \nonumber
& &  \quad \nu' = \frac{d-2d\, h}{d+2}, \qquad z=\frac{\omega\,d}{d+2} w^{\frac{1}{d}+\frac{1}{2}}\sqrt{-f'(0)f'(1)^2}.
\end{eqnarray}
Again, the solution can be matched to the WKB expansion.  
For fixed $w$ since $\omega\rightarrow \infty$, the argument of the Bessel function $z$ is large and therefore we 
get 
\begin{eqnarray}
\lim_{z\rightarrow \infty}   \phi^{(0)}_{\delta=1}(w)   &= &
 B' w^{\frac{d-2}{4d} } \cos \Big(  \frac{\omega d}{ d+ 2} w^{\frac{1}{d}+\frac{1}{2} }\sqrt{-f'(0) f'(1)^2}  - \frac{\pi}{4} ( 2\nu' + 1)  \Big)  \\ \nonumber
 && \qquad +   C' w^{\frac{d-1}{4d} } \sin \Big(  \frac{\omega d}{d+ 2} w^{\frac{1}{d} + \frac{1}{2} }\sqrt{-f'(0)f'(1)^2}  - \frac{\pi}{4} ( 2\nu' + 1)  \Big)
\end{eqnarray}
We  can obtain  $B', C'$   in terms  $A^{(0)\prime}$ and importantly,  they depend on $\omega$ in a common phase
due to the WKB transport of the solution from the horizon to the boundary region. 
However, now the expansion for small $w$ yields
\begin{equation}
\lim_{w\rightarrow 0 }
  \phi^{(0)}_{\delta=1}(w) \approx B^{(0)'}  \omega^{\frac{2dh}{ d+2} } w^h +  C^{(0)'}
   \omega^{\frac{2 ( d- h) }{d+2} } w^{1-h}.
\end{equation}
Here we have extracted the frequency dependence and  $ B^{(0)'}  ,  C^{(0)'} $ are constants independent of the frequency except through a common phase resulting from the WKB transport. 
Therefore the retarded Greens function in the WKB approximation for large frequencies in the light-like limit is given by 
\begin{eqnarray}
  \label{retardplanar}
\lim_{\omega \rightarrow \infty} G^R( \omega, |\vec k| = \omega)\approx \frac{  \omega^{\frac{2dh}{ d+2} } }{ \omega^{\frac{2 ( d- h) }{d+ 2} }}
= w^{ \frac{2}{d+2}  ( 2 \Delta - d) } .
\end{eqnarray}

Our analysis shows that the for planar black holes the retarded correlator  in the light-like limit at large frequencies scales
with a different exponent compared to the generic expected exponent of $(2\Delta -d)$. 
The reason this happened is because, the WKB transports the solution in the near horizon geometry by a phase which can be matched to the near boundary solution. The relevant Greens function then just 
depends on the order of the Bessel functions with order  $\nu$ in (\ref{bessel1}) for generic momenta and $\nu'$ in (\ref{bessel2}) for 
light-like momenta. 

It will be interesting to find if the scaling behaviour 
is independent of the charge of planar black 
hole or if  the blackening function  is a perturbative solution of a higher derivative  theory. 
Note that the  scaling behaviour just depends on the expansion of the potential term in the Schrödinger form of 
the  differential equation of the minimally coupled scalar at boundary in (\ref{largefreqeq}), (\ref{Qexp})
 which dictates the order of the 
Bessel function at the boundary.  Our preliminary calculations indicate that the scaling behaviour is independent 
of the charge as long as $d>4$.

\section{Black holes with spherical and hyperbolic horizons}
\label{sec-sphere}

In this section, we  evaluate the retarded correlator in the light-like limit in black holes with
spherical and hyperbolic horizons. 
We will generalise the WKB approximation developed in the previous section of black hole with planar horizons. 
We consider the more general black hole  metric  \cite{Birmingham:1998nr} \footnote{The $AdS/CFT$ correspondence 
for the $\kappa = -1$ case was discussed in detail by \cite{Emparan:1999gf}.}
\begin{eqnarray}
  \label{eq:BH-metric}
  \dd s^2 &=& \frac{r^2}{R^2} \left(-f(r) \dd t^2\right) + \frac{R^2}{r^2} \frac{1}{f(r)}\dd r^2 + \frac{r^2}{R^2} \dd \Sigma_{k,d-1}^2, \\ \nonumber
  f(r) &=& 1 + \kappa \frac{R^2}{r^2} - \mu\frac{R^2}{r^d}
\end{eqnarray}
where $f(r)$ is the dimensionless blackening function that satisfies $\lim_{r\to\infty} f(r) = 1$, and $\dd \Sigma_{\kappa,d-1}$ is the $d-1$ dimensional horizon metric, which is independent of $r$ and $t$. 
The sign of $\kappa $ determines if the black hole is spherical ($\kappa>0$), planar ($\kappa =0$) of hyperbolic ($\kappa <0$). 
The black holes corresponding to these blackening functions are uncharged. 
The corresponding Reissner-N\"{o}rdstorm black holes contain sub-leading terms at large $r$ in the blackening function $f(r)$, 
 that is terms which depend on the radial coordinate as  $r^{-{2d - 2} }$  and 
 proportional to the charge squared. They are sub-leading for $d>2$ \cite{Chamblin:1999hg,Mann:1997iz}. 
The analysis we develop here applies to all these black holes. 

 The temperature of the black hole can be expressed using the dimensionless blackening function as 
\begin{equation}
  \frac{(r^\star)^2}{R^2}|f'(r^\star)| = \frac{4 \pi}{\beta}.
\end{equation}
The equation of motion of a minimally coupled scalar $f(t,\vec x, r)$ with  mass $m$ nicely splits like
\begin{equation}
  \mathcal D_r f + \mathcal D_t f + \Delta_{\Sigma_\kappa } f - m^2 f = 0,
\end{equation}
where 
\begin{eqnarray}
\mathcal D_r  f = \frac{1}{r^{d-1} R^2  } \partial_r \Big ( r^{d+1}  f(r) \partial_r  f\Big) , \quad 
\mathcal D_t f =  -\frac{R^2}{r^2 f( r) } \partial_t^2 f,  \quad
\mathcal D_{\Sigma_\kappa} f  = \frac{R^2}{r^2} \Delta_{\Sigma_\kappa}   f 
\end{eqnarray}
and $\Delta_{\Sigma_{\kappa }}$ the Laplacian  on the $d-1$  space $\Sigma_{\kappa}$ described the  boundary 
metric $g_{ij}$  parameterised by coordinates $\vec x$. 
We use the ansatz 
\begin{eqnarray}
f(t, \vec x, r) = \exp( - i \Omega t) \phi(\vec x) \psi(r), 
\end{eqnarray}
where $\phi(\vec x)$ are eigenfunctions of  the Laplacian  $\Delta_{\Sigma_{\kappa}}$ with eigenvalues $\hat p( \lambda)$, 
therefore we have 
\begin{eqnarray}
\mathcal D_{\Sigma_\kappa} \phi = \frac{R^2}{r^2}\hat p(\lambda) \phi,  
\end{eqnarray}
To be explicit 
\begin{eqnarray} \label{defplambda}
\hat p( \lambda) &=&   - \vec K^2  , \qquad  \hbox{for}\;  \kappa =0,   \; \vec k  \in \mathbb{R}^{d-1} , \\ \nonumber
&=& - \frac{1}{R^2}
\Big( \lambda^2  + \big( \frac{d-2}{2} \big)^2 \Big) , \qquad   \hbox{for}\; \kappa  = -1,  \lambda \geq 0, \\ \nonumber
& =& -\frac{1}{R^2}  l ( l + d-2) , \qquad \hbox{for}\; \kappa = 1, l = 0, 1, 2, \cdots
\end{eqnarray}
Substituting the ansatz 
The equation of motion for the radial part may be expressed in a new coordinate 
\begin{eqnarray}
\rho=\frac{(r^\star )}{r}
\end{eqnarray}
and takes the form

\begin{equation} \label{basicode}
  \psi''(\rho)+\left(\frac{1-d}{\rho}+\frac{f'(\rho)}{f(\rho)}\right)\psi'(\rho)+\left[ \frac{ f'(1)^2 }{4 } \Big( \frac{\omega^2+f(\rho)p(\lambda)}{f(\rho)^2}  \Big)
  - \frac{d^2 h(h-1)}{\rho^2 f(\rho)}\right] \psi(\rho) = 0
\end{equation}
where 
\begin{eqnarray}
\omega = \frac{\Omega}{2\pi T}, \qquad  p(\lambda)   =  \frac{ \hat p(\lambda) }{ 4\pi^2 T^2} . 
\end{eqnarray}
This equation may be brought into normal form via the transformation 
\begin{equation} 
  \psi(\rho) = \frac{\rho^{\frac{d-1}{2}}}{\sqrt{f(\rho)}} \chi(\rho) \implies \chi''(\rho) + T(\rho) \chi(\rho),
  \label{eq:bh-radial-normal}
\end{equation}
where
\begin{eqnarray}
  \label{eq:potential-schrodinger}
  T(\rho) &=&  \frac{ f'(1)^2 }{4} \Big( \frac{\omega^2+f(\rho)p(\lambda)}{f(\rho)^2}  \Big)
   - \frac{d^2 h(h-1)}{\rho^2 f(\rho)} 
  \\ \nonumber
  & &- \frac{d^2-1}{4 \rho^2} + \frac{\rho f'(\rho)^2+2 f(\rho)((d-1)f'(\rho)-\rho f''(\rho))}{4 \rho f(\rho)^2}.
\end{eqnarray}
Now that the form of the ODE is established, we can begin its analysis. Eventually, we aim to show that the retarded correlator of the field dual to $\phi$ admits a scaling at large $\omega$ in a certain light like limit, that differs from the 
 naive vacuum result. We first analyse the leading order WKB solution at large $\omega$. 
 We then proceed to use the WKB solution which obeys the  ingoing boundary condition 
 at the horizon  and matches the 
 near-boundary expansion, $\rho=0$,  of the ODE. 
 From the latter, the correlator can be extracted via the ratio of the normalisable to the non-normalisable mode.

To define a suitable light-like limit for  all the 3  cases in  (\ref{defplambda}) for which
 $\lambda$ may not correspond to a wave number of a plane wave, we consider the dispersion at the boundary $\rho=0$. In this case, using $f(0) = 1$, the dispersion $\omega^2 + f(w) p(\lambda)$ becomes $\omega^2 + p(\lambda)$. We thus define the light-like limit as 
\begin{equation}
  \label{eq:light-like-condition}
  \omega^2 + p(\lambda) = 0,
\end{equation}
which coincides with the usual definition for plane waves through $p(\lambda) = -\vec k^2$ where $\vec k$ is the dimensionless wave number.

\subsection{WKB approximation}\label{sec:WKB}

We wish to evaluate the retarded correlator at large frequencies both at generic values of the Casimir $p(\lambda)$ 
as well as in the light-like limit  \eqref{eq:light-like-condition}. 
For this we will use the WKB approximation  by taking $\omega\rightarrow \infty$ 
to interpolate or to transport the solution from the near horizon 
region to the near boundary region. 
The horizon $\rho=1$ is a regular singular point of the differential equation (\ref{basicode}), to obtain the retarded solution 
we choose the ingoing boundary condition  at the horizon 
\begin{eqnarray} \label{ingoingg}
\lim_{\rho\rightarrow 1} \psi ( \rho )  =  A  ( 1- \rho)^{ - i \frac{\omega}{2} } 
\end{eqnarray}
We will show that we can choose the WKB solution so that it matches with this ingoing boundary condition
in the inner region and with the near boundary solutions in the near boundary solutions. 

We split the WKB analysis in two cases. We first consider the generic case and then continue with the light like limit, where eq. \eqref{eq:light-like-condition} is satisfied. We define the generic case by $p(\lambda)$ being fixed and  independent of $\omega$ 
\footnote{The analysis can be repeated for a more  general situation in which $p(\lambda) = \delta^2 \omega^2$ 
with $\delta \neq 1$ and $\omega\rightarrow\infty$
as done section \ref{sec:planarbh}.}.  
After we obtain the WKB solutions, in the next subsection we discuss the validity of the approximation and then 
we proceed to match the solution to the near boundary solution.

\paragraph{Generic case:}
If $p(\lambda)$ is generic we choose to approximate 
  $T(\rho)$ asymptotically at large $\omega$ as 
\begin{equation} \label{approxoft}
  T(\rho) \approx \underbrace{ \frac{f'(1)^2 }{ 4  f(\rho)^2}    }_{Q(\rho)} \omega^2.
\end{equation}
The leading order WKB approximation is then 
\begin{equation} \label{wkbapproxg}
  \psi_{\mathrm{WKB}}(\rho) =   A^{(1)}   
    \frac{\rho^{\frac{d-1}{2}}}{f(\rho)^{\frac{1}{2} }    Q(\rho)^{\frac{1}{4} }   }
     \exp\left( i \omega \frac{|f'(1) |}{2}  \int^\rho \frac{dt}{\vert f(t) \vert}    \right) , 
\end{equation}
Near the horizon we have $\rho=1$ and $f(\rho) \approx f'(0)(  \rho -1) $ and $\rho^{\frac{d-1}{2}}$ is regular at $\rho=1$. We therefore obtain the near-horizon expansion 
\begin{equation}
 \lim_{\rho\rightarrow 1}  \psi_{\mathrm{WKB}}(\rho) =  A^{(1)} e^{ - i  \frac{\omega}{2} \log( 1 -\rho) } , 
\end{equation}
which matches the in going solution (\ref{ingoingg}) at the horizon. 
Note that we were not careful will the integration constant, we can choose it to be any point in the domain where the 
WKB approximation is valid,  
  this results in an additional phase proportional to $\omega$ which can be absorbed in the constant. 

We now consider WKB approximation near the boundary $\rho=0$. Here we have
\begin{eqnarray}
\lim_{\rho\rightarrow 0} f(\rho)  &\approx &   1 +  O( \rho^{-m} ) , \qquad m \geq 2
\end{eqnarray}
The condition $m\geq 2$ is obtained from the exact solutions known in (\ref{eq:BH-metric}). 
 The WKB exponent becomes linear in $\rho$  and 
 the near-boundary approximation takes the form 
\begin{equation} \label{wkbgeneral}
  \lim_{\rho\rightarrow 0} \psi_{\mathrm{WKB}}(\rho) =   A^{(0)} 
 \rho^{\frac{d-1}{2} } e^{i \frac{\omega |f'(1)| }{2}  \rho }
\end{equation}
The  phase of the constant $A^{(1)}$, can dependent on $\omega$ linearly. 
This is due to the choice of the lower limit of the integral in the exponent of  (\ref{wkbapproxg}).

\paragraph{Light like case:}
Using the light-like condition \eqref{eq:light-like-condition}, we have 
\begin{equation}
  Q(\rho) = \frac{  f'(1)^2 }{4}  \frac{1-f(\rho) }{f(\rho)^2}.
\end{equation}
With $f(\rho)\approx 0$ near $\rho=1$, it is clear that the near-horizon analysis of the light-like case is completely analogous to the generic case discussed earlier 
The difference to the generic case becomes apparent in the near-boundary region because of the condition $f(0)=1$. The light-like condition is chosen such that this leads to a previously subdominant term in the expansion of $\frac{1-f(\rho)}{f(\rho)^2}$ to carry the relevant contribution. To be precise, 
the dominant term of the Taylor expansion at $\rho=0$ , is cancelled identically in $1-f(\rho)$. Although the cases with $\kappa \in \{-1,0,1\}$
in (\ref{eq:BH-metric})
 only allow for either $\rho^2$ or $\rho^d$ to become dominant, we will for the sake of generality assume that the $n-$th term of the Taylor expansion is dominant. We therefore have 
\begin{equation}
  \lim_{\rho\rightarrow 0} Q(\rho)\approx - \frac{f'(1)^2}{4} \frac{f^{(n)}(0)}{n!}\rho^n.
\end{equation}
The near-boundary approximation is then given by 
\begin{equation} \label{outerwkbg}
  \lim_{\rho\rightarrow 0} \psi_{\mathrm{WKB}}(\rho) =  A^{(0)'} 
   \rho^{\frac{2d-n-2}{4}} e^{i  \omega \frac{2}{2+n} \frac{|f'(1)|}{2} \sqrt{-\frac{f^{(n)}(0)}{n!}} \rho^{\frac{n+2}{2}}}.
\end{equation}
Again, it is important to note that the phase of the  $A^{(0)'} $ can depend on $\omega$ linearly due to the 
choice of the lower limit of the integral in the exponent of the WKB approximation (\ref{wkbapproxg}).

\subsubsection{Turning points and WKB validity}

The WKB approximation is valid if the potential $Q(\rho)$ locally varies slowly compared to the local wave number. The validity criterion is 
\begin{equation}
  \label{eq:WKB-validity}
  \lim_{\omega \to \infty} \left| \frac{Q'(\rho)}{\omega Q(\rho)^{3/2}} \right| \ll 1.
\end{equation}
In the following we aim to analyse the validity of the above WKB approximation and infer matching conditions near the boundary.

\paragraph{Generic case:}
In the generic case, the WKB validity condition \eqref{eq:WKB-validity} evaluates to
\begin{equation}
  \lim_{\omega \to \infty} \left| \frac{f'(\rho)}{\omega f'(1) } \right| \ll 1.
\end{equation}
Thus, there is no turning point and 
WKB approximation is valid throughout the bulk and does not break down  as long as this term,  is the dominant term 
among all the  4 terms in the potential (\ref{eq:potential-schrodinger}). 
Examining the potential we see the we can neglect the other terms as long as
we  are in the regime 
\begin{eqnarray} \label{regime3}
\frac{1}{\omega} \ll \rho\ll1
\end{eqnarray}
This is because the potential is singular at $\rho =0$ for which approximating the potential by just the first term 
using large $\omega$ breaks down. 
Note that very close to the horizon there is a possibility that we need to include the last term in the 
potential  (\ref{eq:potential-schrodinger}), hence the inequality on the right in (\ref{regime3}).

\paragraph{Light like case:}
In the light-like case, the WKB condition reduces to 
\begin{equation}
  \lim_{\omega \to \infty}  \frac{1}{\omega}  \left|  2 f'(1) \frac{(f(\rho)-2)f'(\rho)}{(1-f(\rho))^{3/2}} \right| \ll 1.
\end{equation}
Thus, in this scenario there is a turning point close to the  boundary $\rho=0$. 
To estimate the radial distance  at which the 
validity of the WKB approximation breaks down, we expand the blackening function in $\rho$ up to dominant order $n$ and see when the WKB validity measure reaches $\mathcal O(1)$. This results in the condition 
\begin{equation}
\frac{  \rho^{-\frac{n+2}{2} } } { \omega}   \frac{ 2  ( n+1)!   |f'(1) |}{ | f^{(n)}(0) |^{\frac{1}{2} } } \approx 1
  \implies \rho \sim \mathcal C(n) \omega^{-\frac{2}{n+2}}.
\end{equation}
where $\mathcal C(n) $ is a  constant depending on $n$, which is the leading power $\rho^n$ in the expansion of the 
blackening factor. 
In conclusion, we see that  the boundary layer shrinks to zero as $\omega \to \infty$ which implies the 
scale of the boundary layer is 
\begin{equation}
  \rho_b \sim \omega^{-\frac{2}{2+n}}.
\end{equation}

We have also chosen to approximate the potential in (\ref{eq:potential-schrodinger}), by the first term in the $\omega\rightarrow \infty$ limit. 
We can estimate the radial distance at which this approximation ceases to be valid. 
We need to compare the first term versus terms of order $\rho^{-2}$, this leads to the estimate 
The additional approximation error similar to the generic case does not contribute here. Due to the dominance of the originally subleading term in $1-f(\rho)$, the approximation schematically is 
\begin{equation}
\frac{f'(1)}{4}  \omega^2 \rho^n  f^{(n)}{n!} \approx \frac{1}{\rho^2},  \qquad   \implies \rho \approx \omega^{-\frac{2}{n+2}}.
\end{equation}
where we have neglected sub-leading terms in $\rho$. 
This scale coincides with the scale of the turning point in the WKB approximation. 
Therefore we conclude the in the regime in which the WKB analysis is valid is  given by 
\begin{eqnarray} \label{regime4}
\frac{1}{\omega^{\frac{2}{n+2} } }\ll \rho \ll1 
\end{eqnarray}

\subsubsection{Near boundary solutions}
In the previous sections we have established the WKB solutions at large $\omega$
 and found the domain in the radial co-ordinates in which these solutions are valid 
 both the generic and the light like case. 

The task now is to match the WKB solution to local near-boundary expansions of the full equation. 
This should be possible only if the near boundary solutions  also can be extended in the domain of 
validity of the WKB solutions. 
We again discuss the generic case and the light-like case separately and obtain the 
behaviour of the retarded Greens function for these cases. 

\paragraph{Generic case:}
From (\ref{regime3}), we see that the radial distance at which the WKB analysis breaks down in 
when $\rho \sim \frac{1}{\omega} $. 
This suggests we use the variable 
\begin{eqnarray}
\xi = \omega \rho
\end{eqnarray}
Then the WKB approximation is valid in the domain $\xi \gg1$. 
Near the boundary, keeping the leading terms we approximate the differential equation  for the modified radial mode 
$\chi$  given in 
 (\ref{eq:bh-radial-normal}) as 
\begin{equation} \label{nearboundary2}
  \chi''(\xi) + \left( \frac{1-(2 \Delta-d)^2}{4 \xi^2} + \frac{ f'(1)^2}{4} \right)\chi(\xi) = 0.
\end{equation}
The $\omega$ dependence is implicit in the definition of $\xi$.  Note that, we have  the  leading contribution from the 
first term, together with the second and 3rd terms in the potential (\ref{eq:potential-schrodinger}). 
The solutions are 
\begin{equation} \label{nearboundarygeneric}
  \chi(\xi) = \sqrt{z} \left( B\, J_{\nu}(z) +C\,Y_\nu (z) \right), \quad \nu = \frac{2 \Delta - d}{2},\quad z = 
   \frac{|f'(1)|}{2}  \xi.
\end{equation}
To show that this solution can be matched to the WKB expansion, we expand the solution in the matching region at large $\xi$. The Bessel functions asymptotically behave like 
\begin{equation}
  J_\nu(z) \sim \sqrt{\frac{2}{\pi z}} \cos\left( z-\frac{\pi (1+2\nu)}{4 } \right),\quad Y_\nu(z) \sim \sqrt{\frac{2}{\pi z}} \sin\left( z-\frac{\pi (1+2\nu)}{4} \right). 
\end{equation}
Substituting this  behaviour in the relation of the radial wave function $\psi( \rho)$ defined in (\ref{eq:bh-radial-normal}), 
we obtain 
\begin{eqnarray}
\lim_{\xi\rightarrow \infty} \psi ^{(0)} ( \rho) &=& B \rho^{\frac{d-1 }{2}} 
  \cos\left(  \omega  \frac{ |f'(1)| }{2} \rho  -\frac{\pi (1+2\nu)}{4 } \right) \\ \nonumber
 & &  \qquad\qquad + C \rho^{\frac{d-1}{2}} 
  \sin\left(  \omega  \frac{ |f'(1)|}{2} \rho  -\frac{\pi (1+2\nu)}{4 } \right). 
\end{eqnarray}
Comparing this equation with (\ref{wkbgeneral}), it is evident that we can choose constants $B, C$ such that it 
we can match the inner solution of the near boundary differential equation (\ref{nearboundary2}) to agree with the outer solution 
of the WKB approximation.
Furthermore note that though $A^{(0)}$ may have frequency 
dependence, it is through its phase and it is linear in $\omega$. 
Therefore, matching solving for $B, C$ in terms of $A^{(0)}$, we see that both these constants have the 
same dependence in $\omega$ through the phase.

 Therefore, the retarded two-point function depends on $\omega$ only due to the near-boundary expansion of the Bessel functions. These  expansions are 
\begin{equation}
  \begin{split}
    \lim_{z\rightarrow 0} J_\nu(z) &= \frac{2^{-\nu}  z^{\nu}}{\GA{\nu+1}}\\
    \lim_{z\rightarrow 0 } Y_\nu(z) &= -\frac{1}{\pi}\left( 2^{-\nu} z^\nu \cos \pi \nu \GA{-\nu} + 2^\nu z^{-\nu} \GA{\nu} \right).
  \end{split}
\end{equation}
With $\xi = \omega \rho$  and substituting these expansions in  (\ref{nearboundarygeneric}), 
we obtain 
\begin{eqnarray}
 \lim_{\rho\rightarrow 0} \psi ^{(0)} ( \rho) &=& B^{(0)} \rho^{\Delta} \omega^{\frac{2\Delta - d +1}{2} }
 + C^{(0)} \rho^{d-\Delta} \omega^{ -\frac{ 2\Delta - d - 1}{2} }.
\end{eqnarray}
Here we have absorbed all frequency independent constants as well as the common phase due to the 
WKB transport into the constants  $A^{(0)} , B^{(0)}$. 
It then becomes evident that the retarded two point function scales like 
\begin{equation}
 \lim_{\omega\rightarrow \infty}  G(\omega, p(\lambda)) \sim \omega^{2\nu } = \omega^{2 \Delta - d}.
\end{equation}
This is precisely the expected vacuum scaling that is recovered because thermal effects are washed out at large frequencies.

\paragraph{Light like case:}

For the light-like case,  we see  from (\ref{regime4}) that the WKB analysis is valid till the radial distance
 $\rho  \sim \omega^{-\frac{2}{n+2}}$.  Here $n$ is the index of the first non-zero term  of the Taylor expansion of the blackening function $f(\rho)$ around $\rho =0$. 
 This suggests we use the 
 scaled coordinate 
 \begin{eqnarray}
 \xi = \omega^{\frac{2}{2+n}} \,\rho, 
 \end{eqnarray}
  together with the light-like condition (\ref{eq:light-like-condition})
   to obtain the near boundary approximation of the differential equation (\ref{eq:bh-radial-normal}). 
  Given 
  \begin{eqnarray}
  f(\rho)=1+\frac{f^{(n)}(0)}{n!} \rho^n + \mathcal O(\rho^{n+1}), 
  \end{eqnarray}
   the ODE for the modified radial mode $\chi(\xi)$ in the limit $\omega \to \infty$ is given by 
\begin{equation}
  \chi''(\xi) + \left( \frac{1-d^2 (2h-1)^2}{ 4 \xi^2}   - \frac{f'(1)^2  f^{(n)}(0)}{ 4  n!} \xi^n\right) \chi(\xi) = 0.
\end{equation}
Here again we have kept the leading contribution from the first term, the second and the third term in the potential 
(\ref{eq:potential-schrodinger}). 
The solutions to this equation are modified Bessel functions 
\begin{equation} \label{solgouter}
  \chi(\xi) = \sqrt{\xi} \left( B'\,J_{\nu'}(z) + C'\, Y_{\nu'}(z) \right), \quad \nu' = \frac{2 \Delta - d}{n+2}, \quad z = \frac{2}{2+n}\frac{|f'(1)|}{2}\sqrt{\frac{ - f^{(n)}(0)}{n!}} \xi^{\frac{2+n}{2}}.
\end{equation}
For $\xi$ fixed, together with large  $\omega$,  using the asymptotic expansions of  the Bessel functions we obtain 
\begin{equation}
\lim_{z \rightarrow \infty} \psi^{(0)}(\rho )  =   \rho^{\frac{2d-n-2}{4} } \left[B' \cos\Big( z-\frac{\pi (1+2\nu)}{4 } \Big)+ 
C' \sin\Big( z-\frac{\pi (1+2\nu)}{4 } \Big)  \right] . 
\end{equation}
We have used the relation between the radial mode $\psi(\rho)$ and  modfied mode $\chi(\rho)$ in 
(\ref{eq:bh-radial-normal}).  Comparing the behaviour of the WKB approximation near the boundary in 
(\ref{outerwkbg}) and  that using the definition of $z$ in (\ref{solgouter}),  we see  the solution to the near boundary differential equation 
can be matched to the WKB expansion. 
The constants $B', C'$ do not have any frequency dependence except though an overall phase which can occur 
from $A^{(0) \prime }$ of the WKB solution. 
Because the WKB approximation transports $\omega$ as a phase from the horizon to the boundary, any $\omega$ dependence that is acquired by the matching procedure drops out in the ratio of the normalisable to non-normalisable modes when matching against the purely ingoing solution. How the correlator depends on $\omega$ is then purely determined by the near-boundary expansion of the Bessel functions.

Let us finally examine  the solutions   (\ref{solgouter}) near the boundary. 
We can use the expansion of the Bessel functions in  to show  the the radial mode behaves as 
\begin{equation}
\lim_{\rho\rightarrow 0}\psi^{(0)}(\rho) = A^{(0) \prime} \omega^{\frac{ 2 \Delta - d +1}{n+2} } \rho^{\Delta}
+ B^{(0) \prime} \omega^{ - \frac{ 2 \Delta - d -1}{n+2}} \rho^{d-\Delta} .
\end{equation}
The constants $A^{(0)\prime}, B^{(0)\prime}$ do not depend on the frequency except through an overall common 
phase which can occur due to the WKB transport. 
The ratio of the normalisable and non-normalisable mode then reduces to 
\begin{equation}
  \lim_{\omega\rightarrow \infty} 
  G(\omega,p(\lambda)=-\omega^2) \approx \omega^{2\nu} = \omega^{\frac{2}{2+n}(2\Delta-d)}.
\end{equation}

We may now connect this analysis to particular choices of topological black holes in $AdS_{d+1}$. In the case of spherical or hyperbolic black holes, we have $\kappa = \pm 1$. Whenever $\kappa\neq 0$, the dominant order beyond $n=0$ is 2. Therefore, the light-like retarded two point function in these geometries scales as $\omega^{\frac{2 \Delta - d}{2}}$. In the planar case the curvature parameter $\kappa=0$. Then, the dominant order is $d$. With $n=d$ one then obtains the scaling $
\omega^{\frac{2}{2+d}}$ which agrees with the analysis done in  section \ref{sec:planarbh}  exclusively for the 
case of the planar black hole. 

Furthermore, note that this analysis holds true as long as  the last term in the potential (\ref{eq:potential-schrodinger})
can be neglected in both the 
WKB regime as well as the near boundary regime  for the light-like limit.
We have verified that this is indeed the case for Schwarzschild black holes with planar, spherical as well as 
hyperbolic horizons. 
For charged black holes  we have verified that for  $\kappa = \pm 1$ and $d\geq 3$, the presence of charge does not affect the  and indeed the last term can be neglected.  However for planar black holes the light-like limit analysis is not affected 
only for $d>4$. 
It will be interesting to study the light-limit of the retarded Greens function for charged black holes in detail.

\subsection{Exact result for the hyperbolic black hole}

As a model that is solvable in general dimensions, we will now consider the minimally coupled scalar in the background of a hyperbolic black hole in $AdS_{d+1}$. The metric is given by 
\begin{equation}
  \dd s^2= \frac{R^2}{z^2} \left( -\left( 1- \frac{z^2}{R^2} \right)\dd t^2 + \frac{1}{1-\frac{z^2}{R^2}} \dd z^2 + R^2 \left(\dd u^2 + \sinh^2 u\, \dd \Omega_{d-2}^2\right) \right).
\end{equation}
Here we have set $\kappa =-1$  in (\ref{eq:BH-metric}) and use the coordinate $z = \frac{R^2}{r}$ and explicitly 
written down the metric on the surface $\Sigma_{\kappa, d-1}$. 
The temperature of this black hole is given by 
\begin{eqnarray}
T = \frac{1}{2\pi R} 
\end{eqnarray}
We assume that the Laplacian of $AdS_{d-1}$ acts on its eigenfunctions $\chi(\vec x)$ as 
\begin{equation}
  \Delta \chi(\vec x) = -\left(\lambda^2 +\left(\frac{d-2}{2}\right)^2\right) \chi(\vec x).
\end{equation}
With the Ansatz 
\begin{eqnarray}
\phi(t,z,\vec x) = e^{-i  \frac{ \omega}{R}  t} \chi(\vec x) \psi(z)
\end{eqnarray}
for the bulk scalar $\phi$ we obtain the equation of motion in the variable $w=\frac{z^2}{R^2}$
{\small \begin{equation}
  \psi''(w)+\left( \frac{1}{w-1}+\frac{1-\frac{d}{2}}{w}  \right) \psi'(w) + \left( \frac{(d-2)^2}{16 w (w-1)} + \frac{\Delta(\Delta-d)}{4 w^2 (w-1)} + \frac{ \omega^2 + (w-1)\lambda^2}{4 w (w-1)^2} \right) \psi(w)=0.
\end{equation} }
The general solution to this equation which obeys ingoing boundary conditions at the horizon is given by 
{\small
\begin{eqnarray}
\psi( w) &=&  A^{(1)}  w^{\frac{\Delta}{2}}  (1-w)^{-\frac{i  \omega}{2}}   \\ \nonumber
&& \times\; \GH{\frac{-d+2+2 \Delta+2 i (\lambda- \omega)}{4}}{\frac{-d+2+2 \Delta - 2 i (\lambda+ \omega)}{4}}
{ 1- i \omega }{1-w} 
\end{eqnarray}
}
We can use the connection formula for the hypergeometric function and obtain the behaviour of the solution 
at the boundary $w=1$, from which we obtain the following retarded Greens function
{\small \begin{eqnarray} \label{hyperbolgreen}
 G_R(\omega, \lambda) = ( 2\Delta - d) \frac{R^{d-1}}{ 16\pi G_N}  ( 2\pi T)^{2\Delta - d} 
  \frac{\GA{\frac{2+d-2\Delta}{2}}\GA{\frac{-d+2+2 \Delta+2 i (\lambda- \omega)}{4}}\GA{\frac{-d+2+2 \Delta - 2 i (\lambda+ \omega)}{4}}}{\GA{\frac{2-d+2\Delta}{2}}\GA{\frac{d+2-2 \Delta+2 i (\lambda- \omega)}{4}}\GA{\frac{d+2-2 \Delta - 2 i (\lambda+ \omega)}{4}}}. \nonumber
  \\
\end{eqnarray}}
Observe that this correlator coincides with the retarded Greens function in the BTZ black hole evaluated in 
(\ref{eq:BTZ-correlator-holographic}) once we substitute 
$d =2$ and identify  $\lambda $ with $k$. 
This is expected since the metric of the hyperbolic black hole coincides with the BTZ black hole for $d=2$. 
It is obtain the limiting behaviour of the retarded Greens function in (\ref{hyperbolgreen}) for large frequencies. 
For a fixed $\lambda$ this correlator exhibits the usual vacuum scaling 
\begin{equation}
  G_R(\omega, \lambda) \sim \omega^{2\Delta - d}.
\end{equation}
However, when we first set $\lambda = \pm \, \omega$, the large $\omega$ scaling becomes 
\begin{equation}
  G_R(\omega, \pm  \omega) \sim \omega^{\Delta-\frac{d}{2}}.
\end{equation}
We see that these results coincide with that obtained in  \ref{sec:WKB} for black holes with hyperbolic horizons. 
This exact calculation  therefore serves as a check on the WKB approximation developed in section \ref{sec:WKB}. 

One  point to note is that for the hyperbolic black hole, we can  study the correlator  in the 
dual field theory exactly just as we did in the BTZ case.  This is because the thermal correlator on 
the dual field theory which is on $S^1\times H_{d-1}$ where $H_{d-1}$ is the $d-1$ hyperbolic space can be 
obtained from a conformal transformation of the plane. 
The study of the OPE expansion of this correlator 
 can possibly lead to the identification of the OPE coefficients which contribute to the 
sub-leading behaviour of the Greens function in the light like limit. 
We hope to return to this question in the future.

\section{Light-like limit for retarded stress tensor correlators}
\label{stress}

Properties of 
stress tensor correlators represent universal information  for any conformal field theory. 
The light-like limit of retarded Greens function we are studying in this paper has certainly not been investigated for  
the stress tensor correlator. 
It is  useful to obtain the scaling behaviour of these correlators as this will provide information of the 
 universal properties of the 
OPE expansion  of the stress tensor at least of theories which admit a holographic dual. 
With these motivations,  we study the retarded Greens function for the stress tensor in  its three channels, 
the scalar, the  shear, and the sound channel. 

The equations of motion of the shear and sound mode in the planar black hole background including a Gauss-Bonnet coupling $\lambda$ have been worked out in \cite{Buchel:2009sk}. They are valid for $d\geq 4$. In the limit $\lambda\to 0$, the equations of motion simplify greatly. Like in the  case for the retarded correlator of scalar operators
a general solution of the corresponding ODE in arbitrary dimensions cannot be obtained.
However we can use the WKB method developed in section \ref{sec:WKB} to obtain the scaling behaviour 
in the light-like limit. 
As we have performed the WKB analysis in detail, we will use the observation that once ingoing boundary conditions 
are imposed at the horizon, the WKB approximation transports the wave function to the boundary at most by a phase. 
At the boundary the scaling behaviour just depend on the properties of the Bessel functions. 
We will take this approach to obtaining the scaling exponent, the details  are the same as discussed 
in section \ref{sec:WKB}. 

\paragraph{Scalar mode:}
The equation of motion of the scalar mode  is the same as that of the minimally coupled 
massless scalar, this implies  $\Delta=d$ and the equation is given by 
\begin{equation}
  \phi''(w)+\frac{1}{w-1} \phi'(w) + w^{\frac{2}{d}-2} \frac{\omega^2-(1-w) k^2}{4(w-1)^2} \phi(w) = 0.
\end{equation}
At $k^2 = \omega^2$ this equation of motion becomes
\begin{equation}
  \phi''(w) + \frac{1}{w-1} \phi'(w) + w^{\frac{2}{d}-1} \frac{\omega^2}{4(w-1)^2}\psi(w)=0.
\end{equation}
As discussed in section \ref{sec:WKB}, the WKB analysis is valid in the domain given in 
(\ref{regime4}) with $n=d$, which leads to the fact that the thickness of the  near boundary 
 layer is $w_b = \omega^{-\frac{2d}{d+2}}$ . With this, the near-boundary equation in the scaled variable $\xi = \frac{w}{w_b}$ and at large $\omega$ is 
\begin{equation}
  \phi''(\xi) + \frac{\xi^{\frac{2}{d}-1}}{4}\phi(\xi) = 0.
\end{equation}
The solutions to this are the Bessel functions
\begin{equation}
  \phi(\xi) = \sqrt{\xi} \left( A J_\nu(z) + B Y(\nu)(z) \right), \quad \nu = \frac{d}{d+2}, \quad z = \frac{d}{d+2} \xi^{\frac{d+2}{2d}}.
\end{equation}
Substituting $\xi$ in terms of $w$ and $\omega$ and expanding the Bessel functions for small $w$ then yields the scaling exponent $2\nu = \frac{2d}{d+2}$.  In conclusion we have  
\begin{eqnarray}
\lim_{\omega\rightarrow \infty} G^R( \omega, |\vec k | = \omega) \approx \omega^{\frac{2d}{d+2} }, 
\qquad \hbox{scalar channel of the stress tensor}.
\end{eqnarray}
Note that, if $d\to\infty$, the order of the Bessel functions are an integer. In this case, a  term depending on the 
logarithm of the frequency will appear and a more refined prescription is needed to define the correlator. This will be handled explicitly in section \ref{sec:large-d}.

\paragraph{Shear mode:}

The equation of motion of the shear mode is \footnote{This is equation (4.7) in \cite{Buchel:2009sk}, with the Gauss-Bonnet parameter set to zero. We have also used  the radial co-ordinate defined in (\ref{dcoordinate}). } 
\begin{equation}
  \phi''(w) + \left( \frac{1}{w-1} - \frac{\kappa^2}{\omega^2 - (1-w)\kappa^2} \right)\phi'(w) + w^{\frac{2}{d}-2} \frac{\omega^2 - (1-w)\kappa^2}{4(w-1)^2} \phi(w) = 0.
\end{equation}
Setting $\kappa^2=\omega^2$ one arrives at the equation
\begin{equation}
  \label{shearlight-like}
  \phi''(w)+\frac{1}{w(w-1)}\phi'(w) + w^{\frac{2}{d}-1}\frac{\omega^2}{4(w-1)^2} \phi(w).
\end{equation}
The thickness of the boundary layer is $w_b = \omega^\frac{-2d}{d+2}$. With this, the near-boundary equation in the scaled variable $\xi=\frac{w}{w_b}$ and large $\omega$ is 
\begin{equation}
  \phi''(\xi)-\frac{1}{\xi}\phi'(\xi)+\frac{\xi^{2/d}}{4} \phi(\xi) = 0.
\end{equation}
This ODE is solved by Bessel functions 
\begin{equation}
  \phi(\xi) = \xi \left(A\, J_\nu(z) + B\, Y_{\nu}(z) \right), \quad \nu = \frac{2d}{d+2}, \quad z=\frac{d}{d+2} \xi^{\frac{d+2}{2d}}.
\end{equation}
Substituting $\xi$ in terms of $w$ and $\omega$ and expanding for small $w$ then yields the scaling exponent $2 \nu = \frac{4 d}{d+2}$.  Therefore we obtain 
\begin{eqnarray}
\lim_{\omega\rightarrow \infty} G^R( \omega, |\vec k | = \omega) \approx \omega^{\frac{4d}{d+2} }, 
\qquad \hbox{shear  channel of the stress tensor}.
\end{eqnarray}
Note that the order of the Bessel functions becomes an integer at $d=2$ and $d\to \infty$. The case $d=2$ is excluded due to the validity condition of the stress energy tensor decomposition of $d\geq 4$. In the case of $d\to\infty$ a 
term  depending on the logarithm of the frequency will appear and the prescription to define the correlator needs to be refined. This will be done explicitly in section \ref{sec:large-d}.

\paragraph{Sound mode:}
The equation of motion of the sound mode is \footnote{This is equation (4.11) in \cite{Buchel:2009sk}, the coefficients 
of this differential equation are long expresssions. Once we set the Gauss-Bonnet parameter to zero and take 
 the light-like momenta, the equation simplifies.}
\begin{equation}
  \begin{split}
  \phi''(w) &+ \left( \frac{1}{w-1} - \frac{2(d-2)\kappa^2}{((2+d(w-2)-2w)\kappa^2+2(d-1)\omega^2)} \right)\phi'(w)\\
  &+\left(w^{\frac{2}{d}-2}\frac{\omega^2-(1-w)\kappa^2}{4(w-1)^2}+\frac{1}{2(w-1)}\frac{2(d-2)\kappa^2}{((2+d(w-2)-2w)\kappa^2+2(d-1)\omega^2)}\right)\phi(w) = 0.
  \end{split}
\end{equation}
In the light-like limit this reduces to 
\begin{equation}
  \label{soundlight-like}
  \phi''(w)+\left( \frac{1}{w-1}-\frac{2}{w} \right)\phi'(w) + \frac{-4 (1-w) + w^{2/d}\omega^2}{4(1-w)^2 w} \phi(w) = 0.
\end{equation}
Again, the scale of the boundary layer is $w_b = \omega^{-
\frac{2d}{d+2}}$. The  equation  in terms of $\xi = \frac{w}{w_b}$  at large $\omega$ is 
\begin{equation}
  \phi''(\xi)-\frac{2}{\xi} \phi'(\xi) + \frac{\xi^{\frac{2}{d}-1}}{4} \phi(\xi) = 0.
\end{equation}
This is again a Bessel equation and the solutions are
\begin{equation}
  \phi(\xi) = \xi^{3/2} \left( A\,J_{\nu}(z) + B \, Y_{\nu}(z) \right), \quad \nu=\frac{3d}{d+2}, \quad z=\frac{d}{d+2} \xi^{\frac{d+2}{2d}},
\end{equation}
where $\xi = \frac{w}{w_b}$. Again, the near-boundary expansion yields the scaling exponent $2 \nu = \frac{6d}{d+2}$
which implies that we obtain 
\begin{eqnarray}
\lim_{\omega\rightarrow \infty} G^R( \omega, |\vec k | = \omega) \approx \omega^{\frac{6d}{d+2} }, 
\qquad \hbox{sound channel of the stress tensor}.
\end{eqnarray}
Note that at $d=4$ and $d\to\infty$ the order of the Bessel functions becomes an integer. In these cases, a logarithmic term appears in the expansion of the Bessel function $Y_\nu$. A more refined prescription is then needed to define the correlator. This prescription is used explicitly in section \ref{sec:large-d}. The case of $d=4$ is relevant in the holographic study of $SU(N)$ $\mathcal N=4$ Super-Yang-Mills theory at large $N$.

Recall the scaling exponent of the scalar mode, $\frac{d}{d+2}(2\Delta-d)$. For the scalar mode of the stress-energy tensor $\Delta=d$. Thus, the scaling exponents are 
\begin{equation}
  \alpha_{\mathrm{scalar}} = \frac{2d}{d+2},\quad \alpha_{\mathrm{shear}} = 2\alpha_{\mathrm{scalar}} = \frac{4d}{d+2},\quad \alpha_{\mathrm{sound}}=3\alpha_{\mathrm{scalar}} = \frac{6d}{d+2}.
\end{equation}
If the dimension $d$ is not tuned specifically, there are no logarithmic terms in the correlation function. This happens because the logarithmic term comes with a prefactor of the form $(k^2-\omega^2)^{\Delta/2}$, which vanishes on the light cone. In section \ref{sec:large-d} we compute how the sub-leading behaviour can reintroduce the logarithmic divergence at $d\to \infty$.

\section{Exact retarded correlator at  large $d$ for light-like mometa}
\label{sec:large-d}

The large $d$ limit of gravity has proved to be very useful limit to formulate the dynamics of black hole horizons. 
In the series of works \cite{Emparan:2013moa,Bhattacharyya:2015fdk} it was shown that the black hole horizon in the large $d$ limit can be modelled 
as a membrane  which has properties of a dissipative fluid.  This limit was also studied in the context of the $AdS/CFT$ 
correspondence. 
Properties such a quasi-normal modes and transport coefficients were studied. 
In all these works, the metric perturbations had 
 low frequency and low momentum. One thing that resulted from taking the large $d$ limit was that 
 asymptotic properties of the background metric is not Anti-deSitter \footnote{See \cite{Emparan:2020inr} for a comprehensive review of the large $d$ limit.}.  
 The radial dependence of the fluctuations 
 at generic frequencies and momenta does not have the typical non-normalisable and normalisable modes 
 expected for $AdS/CFT$. 
 
Let us review this briefly. Consider the equation (\ref{eq:EoM-d}),  of the minimally coupled scalar in  the planar $AdS_{d+1}$  black hole
\begin{equation}
  \phi''(w) + \frac{1}{w-1}\phi'(w) + \left( w^{\frac{2}{d}-2} \frac{\omega^2-(1-w)k^2}{4 (w-1)^2} - \frac{h(h-1)}{w^2 (1-w)}\right)\phi(w) = 0,
\end{equation}
Here $(\omega, k = |\vec k| )$ are dimensionless frequency and momenta. 
Taking the large $d$ limit we obtain 
\begin{eqnarray}
 \phi''(w) + \frac{1}{w-1}\phi'(w) + \left(  \frac{\omega^2-(1-w)k^2}{4 w^2  (w-1)^2} - \frac{h(h-1)}{w^2 (1-w)}\right)\phi(w) = 0.
\end{eqnarray}
Note that the behaviour at the boundary $w=0$ which is a regular singular point   of the ODE is altered since the residue 
of the $1/w^2$ coefficient  of the ODE is modified. 
Due to this, the asymptotic  behaviour at the boundary is given by 
\begin{eqnarray}
\lim_{w\rightarrow 0} \phi(w)  =  A w^{\beta_+ } +  B w^{\beta_{-} },   \qquad \beta_\pm \equiv
\frac{1}{2} \Big( 1 \pm \sqrt{ ( 2h -1)^2 - 4 (\omega^2 -k^2) } \Big).
\end{eqnarray}
Note the dependence of the radial dependence of the  behaviour on frequency, which is not consistent with 
the asymptotic  $AdS$ behaviour for a scalar field. 
However, when $\omega = k$, we see that 
\begin{eqnarray}
\beta_{+}|_{\omega = \pm k } = h, \qquad \beta_{-}|_{\omega = \pm  k } = 1-h .
\end{eqnarray}
which is expected $AdS$ behaviour. 
Therefore for light-like momenta it is possible to obtain the holographic Green's function at large $d$. 
 In \cite{David:2022nfn} the $\omega = k =0$ limit was used to obtain holographic thermal one point functions at large $d$. 

In this section, we present an analysis of the minimally coupled scalar at generic $\Delta$ as well as the three modes of the stress energy tensor at large dimension $d\to \infty$ for light-like momenta. 
 In this limit, the equations of motion of the modes become exactly solvable in terms of hypergeometric functions, from which the retarded correlator can be obtained.

\subsection{Minimally coupled scalar at generic $\Delta$}

We now compute the retarded two point function of the minimally coupled scalar with scaling dimension $\Delta$ holographically in the limit of $d\to\infty$ while keeping $h = \frac{\Delta}{d}$ fixed. This limit has been considered in \cite{David:2022nfn}. In the coordinate $w = \frac{z^d}{z_0^d}$, the equation of motion of the radial mode $\phi(z)$ in the planar $AdS_{d+1}$ background is 
\begin{equation} \label{mincoupscaldelta}
  \phi(w) + \frac{1}{w-1}\phi'(w) + \left( w^{\frac{2}{d}-2} \frac{\omega^2-(1-w)k^2}{4 (w-1)^2} - \frac{h(h-1)}{w^2 (1-w)}\right)\phi(w) = 0,
\end{equation} 
where $\omega$ and $k$ are the dimensionless variables in Fourier space. In general, the large $d$ limit of this equation is subtle, because $w^{2/d}$ does not converge uniformly to 1 as $d\to\infty$. However,  as discussed earlier, in  the case $k^2 = \omega^2$ this limit preserves the  asymptotically $AdS$   boundary conditions of a scalar field. 
In this case,  along wih  $\frac{2}{d}\to 0$,  the equation of motion is 
\begin{equation} \label{light-like-larged}
  \phi(w) + \frac{1}{w-1}\phi'(w) + \left( \frac{\omega^2}{4  w (w-1)^2} - \frac{h(h-1)}{w^2 (1-w)}\right)\phi(w) = 0.
\end{equation}
The solution  to this equation are Gauss hypergeometric functions which satisfies the ingoing boundary 
condition at the horizon is given by 
\begin{equation}
  \begin{split}
  \phi(w) =  A (1-w)^{-i \frac{\omega}{2}} w^h \GH{h-\frac{i \omega}{2}}{h-\frac{i \omega}{2}}{1-i \omega }{1-w} .
  \end{split}
\end{equation}
The retarded Green's function can then be obtained by expanding the solution near the boundary and 
taking the ratio of the coefficients of the normalisable mode to the non-normalisable mode. 
We obtain 
\begin{eqnarray} \label{eq:large-d-correlator}
\hat G^R( \omega) \equiv && \lim_{d\rightarrow \infty} \frac{G^R(\omega,  k =  | \omega| ) }{ {\cal K} } =   - \frac{\GA{2-2h}\GA{h-\frac{i \omega}{2}}^2}{\GA{2h} \GA{1-h-\frac{i \omega}{2}}^2},  \\ \nonumber
&& \hbox{where} \; \;{\cal K} =  ( 2 \Delta - d) \frac{ R^{d-1}}{ 16\pi G_N} \Big(\frac{4\pi T}{d} \Big)^{ 2\Delta - d} 
\end{eqnarray}
At large frequencies, the correlator scales as 
\begin{equation}
  \hat G^R(\omega) \sim  \mathcal{C}(h)(- i \omega) ^{4h-2} = \mathcal C(h) (-i \omega)^{2(2h-1)},
\end{equation}
where $\mathcal{C}(h)$ is a constant which depends on $h$. 
We can examine the scaling exponent  of the retarded correlator for the planar black hole 
at arbitrary $d$ evaluated in (\ref{retardplanar}). 
Observe that the exponent coincides with that evaluated above  since 
\begin{equation}
 \lim_{d\to \infty} \frac{2d}{d+2} = 2, \qquad\hbox{and} \quad \Delta = d h.  
 \end{equation}
 This serves as a further check on the scaling exponent of the 
 retarded Green function at light-like momenta. 

Observe that the scaling $\omega^{4h-2}$ is equivalent to $\omega^{2\Delta-d}$ with $d=2$ and the  identifcation
$\Delta|_{2d}  =  2 h$, suggesting a relation of the large $d$ two point function in the light like limit and a BTZ correlator in a kinematic regime in which the momentum $k$ is  held fixed. Indeed, comparing (\ref{eq:large-d-correlator}) and (\ref{eq:BTZ-correlator-holographic}), we find the following 
relation
\begin{eqnarray}
\hat G^R(\omega )  =  {\cal A}  \;G^R( \omega, k =0)  |_{d=2} , 
\qquad {\cal A} = \frac{ 1}{ 4 h} \frac{16\pi G_N}{ R ( 2\pi T)^{2 ( 2h - 1) } }
\end{eqnarray}
In fact we can see this  correspondence with the BTZ correlator, at the level of the ODE of the minimally coupled 
scalar. 
The equation (\ref{eq:d=2}) for the minimally coupled scalar in the BTZ black hole with $k=0$ coincides with the 
(\ref{light-like-larged}), which is  the 
equation of the minimally coupled scalar at 
large $d$ for 
light-like momenta. 
This fact is similar to the  surprising dimensional reduction seen at large $d$ in generic momenta for which the 
relevant geometry is the black hole in 2 dimensions \cite{Soda:1993xc,Emparan:2013xia}. 
This observation is worth investigating further, especially from the dual field theory point of view.

\subsection{Stress energy tensor correlators}

In this section, we evaluate the retarded stress tensor  correlators in all its  3 channels  at large $d$ for light-like limit. 
We obtain exact results for the Greens function in this limit. 

\paragraph{Scalar mode}
The scalar mode of the stress tensor obeys the equation of the massless  minimally coupled scalar. 
Therefore we take $\Delta =d$ in (\ref{mincoupscaldelta}), 
 then taking $d\rightarrow\infty$ with light-ike momenta $\omega^2 -k^2$, we obtain 
\begin{equation} \label{light-like-larged}
  \phi''_\mathrm{scalar}(w) + \frac{1}{w-1}\phi'_\mathrm{scalar}(w) +
   \left( \frac{\omega^2}{4  w (w-1)^2}  \right) \phi_\mathrm{scalar}(w) = 0.
\end{equation}
This equation admits a solution in terms of hypergeometric functions, the solution with infalling
 boundary conditions at the horizon 
is given by 
\begin{equation}
  \phi_\mathrm{scalar}(w) = A\, (1-w)^{-\frac{i \omega}{2}} \GH{-\frac{i \omega}{2}}{-\frac{i \omega}{2}}{1-i \omega}{1-w}.
\end{equation}
Note that the roots of the indicial   equation at the regular singular point $\omega =0$ differ by an integer with the 
one of the roots vanishing.  The second root is unity,  and in terms of the conventional radial co-ordinate which is  related
by 
\begin{equation} \label{defradial}
w= \frac{R^{2d}}{ z_0 r^d },  \qquad \qquad z_0 = \frac{\beta d}{4\pi}, 
\end{equation}
the root is  $\Delta_{\rm scalar}  = d$. 
When the roots differ by an integer, we use the more refined definition of the retarded Greens function 
put forward in \cite{Son:2002sd}. 
It is given by 
\begin{eqnarray} \label{refinedgreen}
G^R( \omega,  k = | \omega| )_{\rm scalar}  = \lim_{d\rightarrow\infty}  \frac{R^{d-1}}{16\pi G_N} \frac{1}{R^{2 \Delta} }   \Big[\lim_{r\to \infty} r^{\Delta+1} \partial_r \log \phi_\mathrm{scalar}(r)\Big],
\end{eqnarray}
where the square brackets indicate that we need to take the finite part in the limit. 
Here $\Delta = \Delta_{\rm scalar} = d$.  The divergences cancel by 
appropriate boundary terms. 
Observe that this definition is dimensionally consistent. 
Evaluating the Green's function using (\ref{refinedgreen}), we obtain
\begin{eqnarray}
\hat G^{R} ( \omega)_{\rm scalar} &=&  
\lim_{d\rightarrow\infty}  - \frac{16\pi G_N}{ R^{d-1}}  \; \frac{z_0^d}{d}  \;   G^R( \omega,  k = \pm \omega )_{\rm scalar} , 
\\ \nonumber
&=&  p_\mathrm{scalar}(\omega) - \frac{\omega^2}{2}\left[ \gamma_E + \psi\Big( -\frac{i \omega}{2} \Big) \right], 
\end{eqnarray}
where
\begin{eqnarray}
  p_\mathrm{scalar}(\omega) = - \frac{i \omega}{2}.  \\ \nonumber
\end{eqnarray}
At large $\omega$, the real and imaginary part of the  correlation function are 
\begin{equation}
  \lim_{\omega\rightarrow\infty} {\rm Im}  \; \hat G^R (\omega)_{\rm scalar}  \sim \omega^2, \qquad\quad
   \lim_{\omega\rightarrow\infty}{\rm Re}  \; \hat G^R(\omega)_{\rm scalar} \sim \omega^2 \log \omega.
\end{equation}

\paragraph{Shear mode}

To solve for the shear mode at large $d$ and for light-like momenta, we examine the equation (\ref{shearlight-like}) and 
take $d\rightarrow\infty$, 
\begin{equation}
  \label{eq:eom-shear-large-d}
  \phi''(w)_{\rm shear} +\frac{1}{w(w-1)}\phi'(w)_{\rm shear} + \frac{\omega^2}{4 w (w-1)^2} \phi(w)_{\rm shear} =0.
\end{equation}
This equation admits a solution in terms of hypergeometric functions, the solution with ingoing boundary 
conditions at the horizon is given by 
\begin{equation}
  \phi_\mathrm{shear} = A\,(1-w)^{-\frac{i \omega}{2}}\GH{-1-\frac{i \omega}{2}}{-\frac{i \omega}{2}}{1-i \omega}{1-w}.
\end{equation}
In terms of the radial co-ordinate defined in (\ref{defradial}), 
the roots of the indicial equation of the ODE  are $0$ and $ 2d$, which results in the effective scaling dimension
$\Delta_{\rm shear} = 2d$. 
We use the prescription given in (\ref{refinedgreen}) to obtain the retarded Greens function, we just need to replace 
$\Delta $ on the LHS by $\Delta_{\rm shear}$, 
which leads to 
\begin{eqnarray}
G^R( \omega,  k = \pm \omega )_{\rm shear}  = \lim_{d\rightarrow\infty}  -  \frac{R^{d-1}}{16\pi G_N} \frac{d}{z_0^{2d}  }   \Big[\lim_{w\to 0}  w^{-1} \partial_w \log \phi_\mathrm{shear}(w)\Big],
\end{eqnarray}
Evaluating the finite term, we obtain 
\begin{eqnarray}
\hat G^{R} ( \omega)_{\rm shear} &=&  
\lim_{d\rightarrow\infty}  - \frac{16\pi G_N}{ R^{d-1}}  \; \frac{z_0^{2 d} }{d}  \;   G^R( \omega,  k = |\omega |)_{\rm shear} , 
\\ \nonumber
&=&   p_\mathrm{shear}(\omega) - \frac{1}{8}\omega^2(4+\omega^2)\left[ \gamma_E + \psi\Big( 1-\frac{i \omega}{2} \Big) \right],
\end{eqnarray}
with 
\begin{eqnarray}
  p_\mathrm{shear}(\omega) =  \frac{\omega}{8} \Big(  4i  \omega + 2 \omega +  i \omega^2 \Big).
\end{eqnarray}
The large frequency behaviour of the retarded Green's function in the shear channel is given by 
\begin{eqnarray}
 \lim_{\omega\rightarrow\infty} {\rm Im}  \; \hat G^R (\omega)_{\rm shear}  \sim \omega^4, \qquad\quad
   \lim_{\omega\rightarrow\infty}{\rm Re}  \; \hat G^R(\omega)_{\rm shear} \sim \omega^4 \log \omega.
\end{eqnarray}

\paragraph{Sound mode}

We follow similar steps to evaluate the retarded Green's function of the sound mode. 
The differential equation determining the Green's function at large $d$ for light-like momenta is obtained 
from (\ref{soundlight-like}) by taking $d\rightarrow\infty$
\begin{equation}
  \phi''(w)_{\rm sound} +\left( \frac{1}{w-1}-\frac{2}{w} \right)\phi'(w)_{\rm sound} + \frac{-4 (1-w) + \omega^2}{4(1-w)^2 w} \phi(w)_{\rm sound} = 0.
\end{equation}
Again the solution to this ODE are hypergeometric functions. 
The solution obeys the ingoing boundary condition at the horizon is given by 
\begin{equation}
  \phi_\mathrm{sound}(w) = A\, (1-w)^{-\frac{i \omega}{2}} \GH{-1-\frac{i \omega}{2}}{-1 - \frac{i \omega}{2}}{1-i\omega}{1-w}.
\end{equation}
The indicial roots at the boundary $w=0$ are $\Delta =0, 3d$ for the radial co-ordinate $r$ defined in (\ref{defradial}). 
This implies the effective conformal dimension of the sound mode is $\Delta_{\rm sound} =3d$. 
From (\ref{refinedgreen}), the prescription to obtain the retarded correlator is given by 
\begin{eqnarray}
G^R( \omega,  k = \pm \omega )_{\rm sound}  = \lim_{d\rightarrow\infty}  -  \frac{R^{d-1}}{16\pi G_N} \frac{d}{z_0^{3d}  }   \Big[\lim_{w\to 0}  w^{-2} \partial_w \log \phi_\mathrm{sound}(w) \Big],
\end{eqnarray}
\begin{eqnarray}
\hat G^{R} ( \omega)_{\rm sound} &=&  
\lim_{d\rightarrow\infty}  - \frac{16\pi G_N}{ R^{d-1}}  \; \frac{z_0^{3 d} }{d}  \;   G^R( \omega,  k = \pm \omega )_{\rm sound} , 
\\ \nonumber
&=&   p_\mathrm{sound}(\omega) - \frac{1}{128} \omega^2(4+\omega^2)^2\left[\gamma_E+\psi\Big(2-\frac{i \omega}{2}\Big) \right],
\end{eqnarray}
with 
\begin{equation}
  p_\mathrm{sound}(\omega) = \frac{1}{512} \Big( -64 + 64 i \omega + 240 \omega^2 + 64 i \omega^3 + 36 \omega^4 +12 i \omega^5  + w^6 \Big) . 
\end{equation}
At large $\omega$, the real and imaginary parts are 
\begin{eqnarray}
 \lim_{\omega\rightarrow\infty} {\rm Im}  \; \hat G^R (\omega)_{\rm shear}  \sim \omega^6, \qquad\quad
   \lim_{\omega\rightarrow\infty}{\rm Re}  \; \hat G^R(\omega)_{\rm shear} \sim \omega^6 \log \omega.
\end{eqnarray}

\section{Conclusions} \label{conclusion}

In this paper we have show that for conformal field theories  that admit a holographic dual, 
the retarded correlator at finite temperature 
admits an anomalous scaling behaviour for light-like momenta at large frequencies. 
For generic momenta, the scaling behaviour at large frequencies is given by 
 $\omega^{ 2\Delta - d}$, where $\Delta$ is the conformal dimension of the operator. 
 The exponent is determined by the identity term in  the OPE expansion of the thermal correlator  and therefore the 
 the thermal effects are washed out and the result is due to 
 by dimensional analysis. 
 
 However as we have shown in this paper for light-like momenta, the scaling exponent is sensitive to the curvature of the 
  horizon.
 For the $AdS$ planar black hole, the retarded correlator scales as $\omega^{\frac{2}{d+2} ( 2\Delta - d) }$, 
 for black holes with spherical and hyperbolic horizons,  it scales as $\omega^{ \frac{2\Delta - d}{2}}$ at large frequencies. 
 This result was established exactly in $d=2$, via numerics in $d=4$ for the planar black hole and using the WKB approximation in general $d$.  The exact calculation for black holes with hyperbolic horizon 
 as well as the  exact analysis at large $d$   also verified the anomalous scaling behaviour. 
 We have evaluated the anomalous scaling exponent for the stress tensor correlator in all its 
 3 channels.

 \paragraph{Discussion}
 It is useful to examine the result from the field theory point of view since our result picks out some 
 universal features of the OPE expansion of the CFT. 
 In view of this,  in appendix  \ref{borelresum}, 
we review the fact that the OPE expansion containing operators
 with $\Delta =J$ was sufficient to reproduce the full retarded correlator  in $d=2$ 
 and therefore also the anomalous scaling. 
 However let us examine other known results in CFT for the retarded correlator to contrast the result 
 for the holographic retarded Greens function. 
 
 For generalised free fields, the retarded correaltor can be obtained from  \cite{Manenti:2019wxs}
  \footnote{We analytically continue the Euclidean correlator in equation (6.3)  of \cite{Manenti:2019wxs}.} and is given by 
 \begin{eqnarray}
 G^R(\omega, \vec k )_{\rm Generalised\;free\;field} = -\frac{\pi^{\frac{d}{2}}  \Gamma( \frac{d}{2} )}{
 2^{2\Delta -d} \Gamma( \Delta) }( \vec k^2 - \omega^2) ^{\Delta - \frac{d}{2}} 
 \end{eqnarray}
 This correlator vanishes for light-like momenta  and does not exhibit any anomalous scaling. 
 Let us examine the large $N$ $O(N)$ model in $d=3$.  The Euclidean correlator is given by 
 \begin{eqnarray}
 G^E(\omega_n, \vec k ) |_{O(N)} = \frac{1}{\omega_n^2 + \vec k^2 + m^2_{\rm th} }
 \end{eqnarray}
 where $\omega_n$ is the Matsubara frequency and $m_{\rm th}$ is the thermal mass. 
 Performing the analytically continuation in (\ref{analyticalcont}) to go over to the retarded Greens function, we see that 
 \begin{eqnarray}
 G^R( \omega, | \vec k | = \omega) |_{O(N)}  =- \frac{1}{m^2_{\rm th} } 
 \end{eqnarray} 
 Therefore, for light-like momenta, the  $O(N)$ model  at large $N$  and finite temperature does not 
 exhibit the anomalous scaling behaviour seen for holographic theories within Einstein gravity. 
 Indeed the critical $O(N)$ model is dual to Vasiliev higher spin gravity. 
 
 Finally let us examine a retarded correlator of the glueball field ${\rm Tr }( F_{\mu\nu}^2)$ evaluated in the 
 free theory, say in $d=4$,  ${\cal N}=4$ Yang-Mills in equation (16) of  \cite{Hartnoll:2005ju}  
 \begin{eqnarray} \label{n=4ym}
 G_R(\omega, k ) |_{F^2} & =&  -\frac{N^2}{\pi^2} (  k^2  - \omega^2)^2 
 \left[ \frac{1}{2} + \Big( i \frac{\pi T}{2k} - \frac{\omega}{ 4k} \Big) \log \frac{ \omega + k }{ \omega - k } 
 + i \frac{\pi T}{k} \log \frac{ \Gamma(   \frac{ -i( \omega+k) }{4\pi T} ) }{ \Gamma(  \frac{-i(\omega-k) }{4\pi T} )}
 \right]   \nonumber \\
 && + \frac{N^2}{\pi^2} \left[ \frac{ 2\pi^2 T^2}{3} ( \omega^2 - k^2)  + \frac{16\pi^2 T^4}{15} 
 + \frac{k^2}{6} \Big( \frac{7k^2}{5} - \omega^2 \Big) \right]
 \end{eqnarray}
 Note the leading contribution at zero temperature here is 
 \begin{eqnarray}
  G_R(\omega, k ) |_{F^2,\;  T=0} & =&  -\frac{N^2}{4 \pi^2}   (  k^2  - \omega^2)^2  \log( k^2 -\omega^2) 
 \end{eqnarray}
 Evaluating the limit of the correlator in (\ref{n=4ym}), we obtain 
 \begin{eqnarray}
  G_R(\omega, k = \omega ) |_{ F^2}  =-  \frac{N^2}{\pi^2} \frac{\omega^4}{60} 
 \end{eqnarray}
 This is a sub-leading correction to the zero temperature result, however its anomalous light-like scaling exponent is not $4/3$ which is the holographic result for an operator of dimension $\Delta =4$ in the geometry of 
 $AdS_5$ planar black hole.

 The comparison of these known results from field theory to the anomalous scaling behaviour of 
 retarded correlators  at  large light-like momenta
  seen in holography show that 
 it is a feature due to the strong coupling behaviour of the theory.  
The dual theory needs to be asymptotically $AdS$ and must exhibit deviations from pure $AdS$ towards the IR region in specific ways, as seen in this paper.  The fact that the scaling behaviour just depends on the dimension of the operator and dimensions of the CFT also highlights its  universal nature.

In \cite{Armas:2025uyv,Armas:2020akg}  the AdS pp-wave geometry  were constructed and studied
 as duals to describe ultra-relativistic fluids in a null limit. 
These geometries are horizonless, nevertheless they deviate from the asymptotic  the $AdS$ geometry at sub-leading orders 
in the radial distance. 
It would be interesting to study the retarded Greens functions  in these backgrounds and obtain its scaling 
behaviour in the light-like limit.

 The OPE in momentum space with careful treatment of the short-distance singularities has been discussed earlier \cite{Bzowski:2019kwd,Bzowski:2014qja}. It is therefore interesting to study the OPE properties in  
 the light-like limit and at finite temperature and derive the behaviour observed in this paper from the CFT in more general cases.

 Lastly, let us point out the simplification seen at large $d$ for correlators with light-like momenta. 
 We saw that  the ODE's determining the 
 correlators  retained the asymptotic  $AdS$ behaviour in the large $d$ limit unlike the case for 
 generic momenta seen in \cite{Emparan:2013moa,Bhattacharyya:2015fdk,Emparan:2020inr}. 
 There was also the surprising dimensional reduction behaviour to the ODE in the BTZ geometry similar 
 to that seen in \cite{Soda:1993xc,Emparan:2013xia}. 
 These features are worth investigating further. 
 
 \acknowledgments
 We thank S. Prem Kumar for correspondence on the results of \cite{Hartnoll:2005ju}. 
 L. S. thanks the warm hospitality the Theoretical Physics II group at Julius-Maximilians-Universität Würzburg during the completion of this work. 
 J.R.D. is partially supported by the ANRF grant ANRF/ARGM/2025/000544/TS. J.R.D and L.S. also thank the organizers and participants of the Advances in Black Hole Theory in IISER Pune and especially Kostas Skenderis for stimulating discussions on the momentum-space OPE. J.R.D. thanks Anshuman Maharana for introducing him to reference \cite{Castells-Tiestos:2022qgu}.

\appendix
\section{Retarded correlators in pure $AdS$}
\label{appena}

In pure $AdS_{d+1}$, the metric is given by the metric \eqref{eq:BH-metric} and $f(r)=1$. With the same conventions as in the previous section, the equations of motion for the radial mode are
\begin{equation}
  \psi''(r) + \frac{d+1}{r} \psi'(r) + \left(R^4\frac{\omega^2 - k^2}{r^4} - \frac{\Delta(\Delta-d)}{r^2} \right)\psi(r) = 0.
\end{equation}
Thus, the difference between pure $AdS$ and a thermal background is that the light-like limit does not admit a WKB expansion at all. There is no sub-leading term that scales with $\omega$ in the limit $\omega^2 - k^2=0$. The general solution is 
\begin{equation}
  \psi(r) = r^{-\frac{d}{2}} \left(A \, I_\nu\left( \frac{R^2}{r} \sqrt{\omega^2 - k^2} \right) + B \, K_\nu\left( \frac{R^2}{r} \sqrt{\omega^2 - k^2} \right) \right),\quad \nu = \frac{2\Delta-d}{2}.
\end{equation}
Imposing regularity at $r=0$ as the boundary condition selects $K_\nu$ as the physical solution. Asymptotically at large $r$ the Bessel function behaves like 
\begin{equation}
  K_\nu(z) \sim 2^{-1-\nu}\GA{-\nu} z^\nu+2^{-1+\nu}\GA{\nu}z^{-\nu},\quad z = \frac{R^2}{r}\sqrt{\omega^2-k^2}.
\end{equation}
This expansion makes the normalisable and non-normalisable modes with exponents $-\Delta$ and $\Delta-d$ apparent. Their ratio results in the retarded two point function 
\begin{equation}
  G_R(\omega,k) \sim 2^{d-2\Delta} R^{4\Delta-2d} \frac{\GA{\frac{d}{2}-\Delta}}{\GA{\Delta-\frac{d}{2}}} (\omega^2-k^2)^{\frac{2\Delta-d}{2}}.
\end{equation}
This zero temperature correlator vanishes identically when $\omega^2 = k^2$. There is no unique scaling behavior that can be singled out, as it solely depends on how this limit is taken.
  
\section{$2d$ Euclidean correlator from  Borel resummation}
\label{borelresum}

In this appendix we show that the Euclidean correlator in $2d$ can be obtained by a Borel resummation of the 
thermal momentum space conformal blocks provided all the OPE coefficients are known. 
This was done earlier in \cite{Barrat:2025nvu}, we review this in more detail and show the occurrence of the anomalous scaling term
at light-like momenta when the Euclidean correlator is continued to the retarded correlator. 
It will also help us isolate the terms in the momentum space OPE that contribute to the anomalous scaling. 
 We will show this explicitly for $\Delta_\phi = 2$, or equivalently $h=\bar h = 1$. The OPE expansion is of the form 
\begin{equation}
  g(\omega_n, k) = \sum_{\Delta=\mathbb N} a_{\Delta,\Delta} f_{\Delta,\Delta}(\omega_n, k),
\end{equation}
where we have taken the limit $J\to\Delta$ in the conformal block $f_{\Delta, J}$ and the OPE coefficient $a_{\Delta, J}$. The conformal blocks in Momentum space have been computed in \cite{Manenti:2019wxs} and are 
\begin{equation}
  \begin{split}
  f_{0,0} &= \frac{\pi}{8} z \bar z \left( \gamma_E-1-\ln 2 + \ln z \right)\\
  f_{\Delta,\Delta} &= 2^{\Delta-3} \pi \GA{\Delta-1} z \bar z\, z^{-\Delta},\quad \Delta>0
  \end{split}
\end{equation}
where $z=k+i \omega_n$. The OPE coefficients are non-zero only if $\Delta \in 2 \mathbb N$ \cite{Barrat:2025nvu} and at $J=\Delta$ in $d=2$ they are 
\begin{equation}
  a_{\Delta, \Delta} = 2(\Delta-1) \zeta(\Delta).
\end{equation}
The correlator may then be expressed as a formal power series of the form 
\begin{equation}
  \label{eq:OPE-not-resummed}
  G_{\mathrm{OPE}}(k,\omega_n) = f_{0,0} + \sum_{\Delta \in 2 \mathbb N} 2^{\Delta-2} \pi \GA{\Delta}\zeta(\Delta) z \bar z \, z^{-\Delta},
\end{equation}
which diverges for generic values of $k$ and $\omega_n$. As in \cite{Barrat:2025nvu} we therefore consider the Borel resummation of \ref{eq:OPE-not-resummed}. Considering the growth of the coefficient in the sum in $\Delta \in \mathbb N$, the Borel transformation is given by 
\begin{equation}
  \mathcal B\left[ G_\mathrm{OPE} \right](t) = f_{0,0} + \sum_{\Delta \in 2 \mathbb N} 2^{\Delta-2} \pi \GA{\Delta} \zeta(\Delta) z \bar z \, z^{-\Delta} \frac{t^\Delta}{\Delta!}.
\end{equation}
We may rewrite this sum using the new summation index $\Delta = 2m$ where $m\in \mathbb N$, resulting in the Borel transform 
\begin{equation}
  \begin{split}
  \mathcal B\left[ G_\mathrm{OPE} \right](t) &= f_{0,0} + \sum_{m =1}^\infty 4^{m-1} \pi \GA{2m} \zeta(2m) z \bar z \, z^{-2m} \frac{t^{2m}}{(2m)!}\\
  &= f_{0,0} + z \bar z \frac{\pi}{8}\left( \ln\GA{1-2 \frac{t}{z}}+\ln\GA{1+2 \frac{t}{z}}\right).
  \end{split}
\end{equation}
The re-summed correlator is then obtained by the inverse Laplace transform 
\begin{equation}
  G_\mathrm{OPE}(z) = \int_\gamma \mathcal B \left[ G_\mathrm{OPE} \right](t) e^{-t} \dd t,
\end{equation}
where $\gamma$ is a suitable integration path, and the integral is understood to be defined on the Riemann surface of the integrand. Since we wish to show that the anomalous light-like scaling can emerge from the OPE, we are not interested in the full non-perturbative structure. For simplicity, we will therefore restrict to the principal sheet in the integration and neglect the contributions from branch cut crossings. This way we can perform a partial integration and neglect the boundary terms coming from the branch cut crossings entirely. Thus, the principal sheet contributions are 
\begin{equation}
  G_\mathrm{OPE}(z) = f_{0,0} + \int_0^\infty \frac{\pi z \bar z}{8}\left( \frac{1}{t} - \frac{2 \pi}{z}\cot \frac{2 \pi t}{z} \right) e^{-t}\dd t.
\end{equation}
For $\Im(z) \neq 0$ this integral converges and can be solved using the Fourier series of the cotangent,
\begin{equation}
  \cot x = - i \left( 1 + 2 \sum_{k=1}^\infty e^{2 i k x} \right).
\end{equation}
Regularising the integral with a lower cutoff of $0<\epsilon\ll 1$ each term in the integral becomes convergent and the integrals may be pulled apart. The first integral is the incomplete $\Gamma$ function
\begin{equation}
  I_1 = \int_\epsilon^\infty \frac{e^{-t}}{t} \dd t = \Gamma(0,\epsilon).
\end{equation}
Next, one must consider 
\begin{equation}
  -\frac{2 \pi}{z} \cot \frac{2 \pi t}{z} = \frac{2 \pi i}{z} + \sum_{k=1}^\infty \frac{4 \pi i}{z} e^{4 \pi i k \frac{t}{z}}.
\end{equation}
This leads to the two additional integrals 
\begin{equation}
  I_2 = \int_\epsilon^\infty \frac{2 \pi i}{z} e^{-t}   \dd t = \frac{2 \pi i}{z} e^{-\epsilon}
\end{equation}
and 
\begin{equation}
  I_3 = \int_\epsilon^\infty  \Big( \sum_{k=1}^\infty  \frac{4 \pi i}{z} e^{4 \pi i k \frac{t}{z}}e^{-t} \Big)  \dd t.
\end{equation}
In the latter integral may now be integrated term by term:
\begin{equation}
  I_3 = \sum_{k=1}^\infty \int_\epsilon^\infty \Big( \frac{4 \pi i}{z} e^{4 \pi i k \frac{t}{z}} e^{-t} \Big) \dd t 
   = \sum_{k=1}^\infty - \frac{4 \pi}{4 \pi k+i z} e^{\left(\frac{4 \pi i k}{z}-1\right) \epsilon} = -\mathrm B_{e^{\frac{4 \pi i \epsilon}{z}}} \left( 1+\frac{i z}{4 \pi} , 0 \right),
\end{equation}
where $\mathrm{B}_z\left( a, b \right)$ is the incomplete beta function. Although there are conditions of convergence on $z$, the beta function provides the necessary analytical continuation for arbitrary z. The incomplete $\Gamma$ function as well as the beta function diverge in the limit $\epsilon \to 0$, which is really the statement that the original integral cannot be separated into sums if $\epsilon=0$. However, their divergences cancel exactly. We have 
\begin{equation}
  \begin{split}
    \Gamma \left( 0,\epsilon \right) &= -\gamma_E - \ln \epsilon + \mathcal O(\epsilon)\\
    \mathrm B_{e^{\frac{4 \pi i \epsilon}{z}}} \left( 1+\frac{i z}{4 \pi} , 0 \right) &= -\gamma_E  - \ln \epsilon - \ln \frac{4 \pi i}{z} - \psi\left(1+\frac{i z}{4 \pi}\right) + \mathcal O (\epsilon).
  \end{split}
\end{equation}
The full integral in the limit $\epsilon \to 0$ is then 
\begin{equation}
  \label{eq:GOPE-missing}
 G_\mathrm{OPE}(z) = f_{0,0} + \frac{\pi z \bar z}{8}\left(I_1 + I_2 + I_3\right) = f_{0,0} + \frac{\pi^2 i \bar z}{4} + \frac{\pi}{8} z \bar z \ln \frac{4 \pi i}{z} + \frac{\pi}{8} z \bar z \psi\left( 1 + \frac{i z}{4 \pi} \right).
\end{equation}
It is important to note that this correlation function cannot yet be complete, because it does not satisfy periodicity and conjugate symmetry in euclidean position space, which manifests as a reality condition of the form
\begin{equation}
  G_\mathrm{OPE}(z) = \bar G_\mathrm{OPE}(\bar z).
\end{equation} 
This can be seen from the Fourier series
\begin{equation}
  \begin{split}
    G_\mathrm{OPE}(\omega_n) &= \int_0^\beta \dd \tau e^{-i \omega_n \tau} G_\mathrm{OPE}(\tau)\\
    &\overset{\mathrm{symmetry}}{=} \int_0^\beta \dd \tau e^{-i \omega_n \tau} \bar G_\mathrm{OPE}(-\tau)\\
    &\overset{\tau'=-\tau}{=} \int_{-\beta}^0 \dd \tau' e^{i \omega_n \tau'} \bar G_\mathrm{OPE}(\tau')\\
    &\overset{\mathrm{periodicity}}{=} \int_0^\beta \dd \tau' e^{i \omega_n \tau'} \bar G_\mathrm{OPE}(\tau')\\
    &= \bar G_\mathrm{OPE}(-\omega_n).
  \end{split}
\end{equation}
Here, we have suppressed the dependence on $k$. However, the symmetry $\omega_n\leftrightarrow-\omega_n$ exactly manifests as a symmetry in $z \leftrightarrow \bar z$ with additional complex conjugation. We restore this symmetry by adding the complex conjugate part of \eqref{eq:GOPE-missing}. The full correlator is then 
\begin{equation}
  \begin{split}
  G(k, \omega_n) &= f_{0,0} + \bar f_{0,0} + \frac{i \pi^2}{4}\left( \bar z - z \right) + \frac{\pi}{8} z \bar z\, \left(\ln \frac{4 \pi i}{z} + \ln \frac{-4 \pi i}{\bar z} \right)\\
  &+ \frac{\pi}{8} z \bar z\, \left(\psi \left( 1+\frac{i z}{4 \pi}\right) + \psi \left( 1-\frac{i \bar z}{4 \pi}\right)\right).
  \end{split}
\end{equation}
This can be simplified if ${\rm Re} (z) \neq 0$ and ${\rm Im }\nleq 0$. The result is 
\begin{equation}
  \begin{split}
  G(k, \omega_n) &= f_{0,0} + \bar f_{0,0} + \frac{i \pi^2}{4}\left( \bar z - z \right) + \frac{\pi}{8} z \bar z\, \left(\ln 16 \pi^2 - \ln z \bar z \right)\\
  &+ \frac{\pi}{8} z \bar z\, \left(\psi \left( 1+\frac{i z}{4 \pi}\right) + \psi \left( 1-\frac{i \bar z}{4 \pi}\right)\right)\\
  &= \frac{i \pi^2}{4}\left( \bar z - z \right) + \frac{\pi}{8} z \bar z\, \left(\ln 16 \pi^2 + 2 \gamma_E-2-\ln 4\right)\\
  &+ \frac{\pi}{8} z \bar z\, \left(\psi \left( 1+\frac{i z}{4 \pi}\right) + \psi \left( 1-\frac{i \bar z}{4 \pi}\right)\right),
  \end{split}
\end{equation}
where in the last line we have used 
\begin{equation}
  f_{0,0}+\bar f_{0,0} = \frac{\pi}{8} z \bar z\, \left( -2+2 \gamma_E-\ln 4 + \ln z \bar z\right),\quad {\rm Re}(z)\leq0, {\rm Im }(z)=0\;\mathrm{excluded}.
\end{equation}
Using $\psi(1-x)=\pi \cot(\pi x) - \frac{1}{x} + \psi(1+x)$, this can be brought into the canonical form
\begin{equation}
  \begin{split}
  G(k,\omega_n) &= \frac{i \pi^2}{4}\left( z-\bar z \right) + \frac{\pi}{8} z \bar z\, \left(\ln 16 \pi^2 + 2 \gamma_E-2-\ln 4\right)\\
  &+ \frac{\pi}{8} z \bar z\, \left(\psi \left( 1-\frac{i z}{4 \pi}\right) + \psi \left( 1+\frac{i \bar z}{4 \pi}\right) + i \pi \left( \coth\frac{z}{4}-\coth \frac{\bar z}{4} \right)\right)
  \end{split}
\end{equation}
With $z=k+i \omega_n$ we eventually obtain 
\begin{equation}
  \begin{split}
  G(k,\omega_n) &= -\frac{\pi^2 \omega_n}{2} + \frac{\pi}{4}(k^2 + \omega_n^2) \left( \ln 2 \pi + \gamma_E - 1 \right)\\
  &+\frac{\pi}{8}(k^2+\omega_n^2) \left( \psi\left( 1 + \frac{\omega_n}{4 \pi} - \frac{i k}{4 \pi} \right) + \psi\left( 1 + \frac{\omega_n}{4 \pi} + \frac{i k}{4 \pi} \right) - \underbrace{\frac{2 \pi \sin \frac{\omega_n}{2}}{\cos \frac{\omega_n}{2} - \cosh \frac{k}{2}}}_{\to 0\; \mathrm{for}\;\omega_n = 2 n \pi}\right).
  \end{split}
\end{equation}
Using the Matsubara frequencies $\omega_n = 2 \pi n$ we see that this coincides with the correlation function obtained directly from the Fourier transform in (\ref{eq:Euclidean-Correlator-h=1}), 
 up to contact terms and overall normalisation. The anomalous term is present in the $z-\bar z$ contribution coming from the integral $I_2$ as well as Digamma function properties in $I_3$. Thus, we have shown that OPE at least in $d=2$ contains the information about the anomalous scaling, and that the term can be made manifest upon Borel resummation of the formal expansion.

\bibliographystyle{JHEP}
 \bibliography{biblio.bib}

@article{Son:2002sd,
    author = "Son, Dam T. and Starinets, Andrei O.",
    title = "{Minkowski space correlators in AdS / CFT correspondence: Recipe and applications}",
    eprint = "hep-th/0205051",
    archivePrefix = "arXiv",
    reportNumber = "INT-PUB-02-34",
    doi = "10.1088/1126-6708/2002/09/042",
    journal = "JHEP",
    volume = "09",
    pages = "042",
    year = "2002"
}

@article{Iqbal:2009fd,
    author = "Iqbal, Nabil and Liu, Hong",
    editor = "Lust, Dieter and Dobrev, Vladimir",
    title = "{Real-time response in AdS/CFT with application to spinors}",
    eprint = "0903.2596",
    archivePrefix = "arXiv",
    primaryClass = "hep-th",
    reportNumber = "MIT-CTP-4022",
    doi = "10.1002/prop.200900057",
    journal = "Fortsch. Phys.",
    volume = "57",
    pages = "367--384",
    year = "2009"
}

@article{Maldacena:1997re,
    author = "Maldacena, Juan Martin",
    title = "{The Large $N$ limit of superconformal field theories and supergravity}",
    eprint = "hep-th/9711200",
    archivePrefix = "arXiv",
    reportNumber = "HUTP-97-A097, HUTP-98-A097",
    doi = "10.4310/ATMP.1998.v2.n2.a1",
    journal = "Adv. Theor. Math. Phys.",
    volume = "2",
    pages = "231--252",
    year = "1998"
}

@article{Brown:1986nw,
    author = "Brown, J. David and Henneaux, M.",
    title = "{Central Charges in the Canonical Realization of Asymptotic Symmetries: An Example from Three-Dimensional Gravity}",
    doi = "10.1007/BF01211590",
    journal = "Commun. Math. Phys.",
    volume = "104",
    pages = "207--226",
    year = "1986"
}

@article{Kovtun:2003wp,
    author = "Kovtun, Pavel and Son, Dam T. and Starinets, Andrei O.",
    title = "{Holography and hydrodynamics: Diffusion on stretched horizons}",
    eprint = "hep-th/0309213",
    archivePrefix = "arXiv",
    reportNumber = "UW-PT-03-21, INT-PUB-03-17",
    doi = "10.1088/1126-6708/2003/10/064",
    journal = "JHEP",
    volume = "10",
    pages = "064",
    year = "2003"
}

@article{Kovtun:2004de,
    author = "Kovtun, P. and Son, Dan T. and Starinets, Andrei O.",
    title = "{Viscosity in strongly interacting quantum field theories from black hole physics}",
    eprint = "hep-th/0405231",
    archivePrefix = "arXiv",
    reportNumber = "INT-PUB-04-09, UW-PT-04-04",
    doi = "10.1103/PhysRevLett.94.111601",
    journal = "Phys. Rev. Lett.",
    volume = "94",
    pages = "111601",
    year = "2005"
}

@inproceedings{Hubeny:2011hd,
    author = "Hubeny, Veronika E. and Minwalla, Shiraz and Rangamani, Mukund",
    title = "{The fluid/gravity correspondence}",
    booktitle = "{Theoretical Advanced Study Institute in Elementary Particle Physics}: {String theory and its Applications: From meV to the Planck Scale}",
    eprint = "1107.5780",
    archivePrefix = "arXiv",
    primaryClass = "hep-th",
    pages = "348--383",
    year = "2012"
}

@article{Hartnoll:2009sz,
    author = "Hartnoll, Sean A.",
    editor = "Uranga, A. M.",
    title = "{Lectures on holographic methods for condensed matter physics}",
    eprint = "0903.3246",
    archivePrefix = "arXiv",
    primaryClass = "hep-th",
    doi = "10.1088/0264-9381/26/22/224002",
    journal = "Class. Quant. Grav.",
    volume = "26",
    pages = "224002",
    year = "2009"
}

@article{Dodelson:2023nnr,
    author = "Dodelson, Matthew and Iossa, Cristoforo and Karlsson, Robin and Lupsasca, Alexandru and Zhiboedov, Alexander",
    title = "{Black hole bulk-cone singularities}",
    eprint = "2310.15236",
    archivePrefix = "arXiv",
    primaryClass = "hep-th",
    reportNumber = "CERN-TH-2023-192",
    doi = "10.1007/JHEP07(2024)046",
    journal = "JHEP",
    volume = "07",
    pages = "046",
    year = "2024"
}

@article{Hartnoll:2005ju,
    author = "Hartnoll, Sean A. and Kumar, S. Prem",
    title = "{AdS black holes and thermal Yang-Mills correlators}",
    eprint = "hep-th/0508092",
    archivePrefix = "arXiv",
    reportNumber = "DAMTP-2005-73",
    doi = "10.1088/1126-6708/2005/12/036",
    journal = "JHEP",
    volume = "12",
    pages = "036",
    year = "2005"
}

@article{Festuccia:2005pi,
    author = "Festuccia, Guido and Liu, Hong",
    title = "{Excursions beyond the horizon: Black hole singularities in Yang-Mills theories. I.}",
    eprint = "hep-th/0506202",
    archivePrefix = "arXiv",
    reportNumber = "MIT-CTP-3641",
    doi = "10.1088/1126-6708/2006/04/044",
    journal = "JHEP",
    volume = "04",
    pages = "044",
    year = "2006"
}

@article{Horowitz:1999jd,
    author = "Horowitz, Gary T. and Hubeny, Veronika E.",
    title = "{Quasinormal modes of AdS black holes and the approach to thermal equilibrium}",
    eprint = "hep-th/9909056",
    archivePrefix = "arXiv",
    reportNumber = "NSF-ITP-99-70",
    doi = "10.1103/PhysRevD.62.024027",
    journal = "Phys. Rev. D",
    volume = "62",
    pages = "024027",
    year = "2000"
}

@article{Berti:2009kk,
    author = "Berti, Emanuele and Cardoso, Vitor and Starinets, Andrei O.",
    title = "{Quasinormal modes of black holes and black branes}",
    eprint = "0905.2975",
    archivePrefix = "arXiv",
    primaryClass = "gr-qc",
    doi = "10.1088/0264-9381/26/16/163001",
    journal = "Class. Quant. Grav.",
    volume = "26",
    pages = "163001",
    year = "2009"
}

@article{Iliesiu:2018fao,
    author = "Iliesiu, Luca and Kolo{\u{g}}lu, Murat and Mahajan, Raghu and Perlmutter, Eric and Simmons-Duffin, David",
    title = "{The Conformal Bootstrap at Finite Temperature}",
    eprint = "1802.10266",
    archivePrefix = "arXiv",
    primaryClass = "hep-th",
    reportNumber = "CALT-TH-2018-013, PUPT-2550",
    doi = "10.1007/JHEP10(2018)070",
    journal = "JHEP",
    volume = "10",
    pages = "070",
    year = "2018"
}

@article{Alday:2020eua,
    author = "Alday, Luis F. and Kologlu, Murat and Zhiboedov, Alexander",
    title = "{Holographic correlators at finite temperature}",
    eprint = "2009.10062",
    archivePrefix = "arXiv",
    primaryClass = "hep-th",
    reportNumber = "CERN-TH-2020-155",
    doi = "10.1007/JHEP06(2021)082",
    journal = "JHEP",
    volume = "06",
    pages = "082",
    year = "2021"
}

@article{Barrat:2025nvu,
    author = "Barrat, Julien and Bozkurt, Deniz N. and Marchetto, Enrico and Miscioscia, Alessio and Pomoni, Elli",
    title = "{The analytic bootstrap at finite temperature}",
    eprint = "2506.06422",
    archivePrefix = "arXiv",
    primaryClass = "hep-th",
    reportNumber = "DESY-25-078",
    month = "6",
    year = "2025"
}

@article{Barrat:2025twb,
    author = "Barrat, Julien and Bozkurt, Deniz N. and Marchetto, Enrico and Miscioscia, Alessio and Pomoni, Elli",
    title = "{Analytic thermal bootstrap meets holography}",
    eprint = "2510.20894",
    archivePrefix = "arXiv",
    primaryClass = "hep-th",
    reportNumber = "DESY-25-139 , YITP-SB-2025-16",
    month = "10",
    year = "2025"
}

@article{Buric:2025anb,
    author = "Buri{\'c}, Ilija and Gusev, Ivan and Parnachev, Andrei",
    title = "{Thermal holographic correlators and KMS condition}",
    eprint = "2505.10277",
    archivePrefix = "arXiv",
    primaryClass = "hep-th",
    doi = "10.1007/JHEP09(2025)053",
    journal = "JHEP",
    volume = "09",
    pages = "053",
    year = "2025"
}

@article{Buric:2025fye,
    author = "Buri{\'c}, Ilija and Gusev, Ivan and Parnachev, Andrei",
    title = "{Holographic correlators from thermal bootstrap}",
    eprint = "2508.08373",
    archivePrefix = "arXiv",
    primaryClass = "hep-th",
    doi = "10.1007/JHEP05(2026)059",
    journal = "JHEP",
    volume = "05",
    pages = "059",
    year = "2026"
}

@article{Petkou:2018ynm,
    author = "Petkou, Anastasios C. and Stergiou, Andreas",
    title = "{Dynamics of Finite-Temperature Conformal Field Theories from Operator Product Expansion Inversion Formulas}",
    eprint = "1806.02340",
    archivePrefix = "arXiv",
    primaryClass = "hep-th",
    reportNumber = "CERN-TH-2018-132",
    doi = "10.1103/PhysRevLett.121.071602",
    journal = "Phys. Rev. Lett.",
    volume = "121",
    number = "7",
    pages = "071602",
    year = "2018"
}

@article{Gobeil:2018fzy,
    author = "Gobeil, Yan and Maloney, Alexander and Ng, Gim Seng and Wu, Jie-qiang",
    title = "{Thermal Conformal Blocks}",
    eprint = "1802.10537",
    archivePrefix = "arXiv",
    primaryClass = "hep-th",
    doi = "10.21468/SciPostPhys.7.2.015",
    journal = "SciPost Phys.",
    volume = "7",
    number = "2",
    pages = "015",
    year = "2019"
}

@article{Dodelson:2022yvn,
    author = "Dodelson, Matthew and Grassi, Alba and Iossa, Cristoforo and Panea Lichtig, Daniel and Zhiboedov, Alexander",
    title = "{Holographic thermal correlators from supersymmetric instantons}",
    eprint = "2206.07720",
    archivePrefix = "arXiv",
    primaryClass = "hep-th",
    reportNumber = "CERN-TH-2022-095",
    doi = "10.21468/SciPostPhys.14.5.116",
    journal = "SciPost Phys.",
    volume = "14",
    pages = "116",
    year = "2023"
}

@article{David:2023uya,
    author = "David, Justin R. and Kumar, Srijan",
    title = "{Thermal one-point functions: CFT{\textquoteright}s with fermions, large d and large spin}",
    eprint = "2307.14847",
    archivePrefix = "arXiv",
    primaryClass = "hep-th",
    doi = "10.1007/JHEP10(2023)143",
    journal = "JHEP",
    volume = "10",
    pages = "143",
    year = "2023"
}

@article{Dodelson:2023vrw,
    author = "Dodelson, Matthew and Iossa, Cristoforo and Karlsson, Robin and Zhiboedov, Alexander",
    title = "{A thermal product formula}",
    eprint = "2304.12339",
    archivePrefix = "arXiv",
    primaryClass = "hep-th",
    reportNumber = "CERN-TH-2023-062",
    doi = "10.1007/JHEP01(2024)036",
    journal = "JHEP",
    volume = "01",
    pages = "036",
    year = "2024"
}

@article{Bhattacharya:2025vyi,
    author = "Bhattacharya, Jyotirmoy and Padhi, Nibedita and Sharma, Aditya and Singha, Sourav",
    title = "{Thermal product formula for shear modes}",
    eprint = "2504.17781",
    archivePrefix = "arXiv",
    primaryClass = "hep-th",
    doi = "10.1007/JHEP08(2025)170",
    journal = "JHEP",
    volume = "08",
    pages = "170",
    year = "2025"
}

@article{Jia:2025jbi,
    author = "Jia, Hewei Frederic and Rangamani, Mukund",
    title = "{Thermal spectral function asymptotics and black hole singularity in holography}",
    eprint = "2512.15114",
    archivePrefix = "arXiv",
    primaryClass = "hep-th",
    month = "12",
    year = "2025"
}

@article{Jia:2026ryl,
    author = "Jia, Hewei Frederic and Rangamani, Mukund",
    title = "{Exact holographic thermal spectral functions: OPE, non-perturbative corrections, and black hole singularity}",
    eprint = "2604.10803",
    archivePrefix = "arXiv",
    primaryClass = "hep-th",
    month = "4",
    year = "2026"
}

@article{Hubeny:2006yu,
    author = "Hubeny, Veronika E and Liu, Hong and Rangamani, Mukund",
    title = "{Bulk-cone singularities {\&} signatures of horizon formation in AdS/CFT}",
    eprint = "hep-th/0610041",
    archivePrefix = "arXiv",
    reportNumber = "DCPT-06-29, MIT-CTP-3775",
    doi = "10.1088/1126-6708/2007/01/009",
    journal = "JHEP",
    volume = "01",
    pages = "009",
    year = "2007"
}

@article{Dodelson:2020lal,
    author = "Dodelson, Matthew and Ooguri, Hirosi",
    title = "{Singularities of thermal correlators at strong coupling}",
    eprint = "2010.09734",
    archivePrefix = "arXiv",
    primaryClass = "hep-th",
    doi = "10.1103/PhysRevD.103.066018",
    journal = "Phys. Rev. D",
    volume = "103",
    number = "6",
    pages = "066018",
    year = "2021"
}

@article{Huang:2022vet,
    author = "Huang, Kuo-Wei and Karlsson, Robin and Parnachev, Andrei and Valach, Samuel",
    title = "{Freedom near lightcone and ANEC saturation}",
    eprint = "2210.16274",
    archivePrefix = "arXiv",
    primaryClass = "hep-th",
    reportNumber = "CERN-TH-2022-173",
    doi = "10.1007/JHEP05(2023)065",
    journal = "JHEP",
    volume = "05",
    pages = "065",
    year = "2023"
}

@article{Esper:2023jeq,
    author = "Esper, Chantelle and Huang, Kuo-Wei and Karlsson, Robin and Parnachev, Andrei and Valach, Samuel",
    title = "{Thermal stress tensor correlators near lightcone and holography}",
    eprint = "2306.00787",
    archivePrefix = "arXiv",
    primaryClass = "hep-th",
    reportNumber = "CERN-TH-2023-091",
    doi = "10.1007/JHEP11(2023)107",
    journal = "JHEP",
    volume = "11",
    pages = "107",
    year = "2023"
}

@article{Caron-Huot:2006pee,
    author = "Caron-Huot, Simon and Kovtun, Pavel and Moore, Guy D. and Starinets, Andrei and Yaffe, Laurence G.",
    title = "{Photon and dilepton production in supersymmetric Yang-Mills plasma}",
    eprint = "hep-th/0607237",
    archivePrefix = "arXiv",
    doi = "10.1088/1126-6708/2006/12/015",
    journal = "JHEP",
    volume = "12",
    pages = "015",
    year = "2006"
}

@article{Castells-Tiestos:2022qgu,
    author = "Castells-Tiestos, Luc{\'\i}a and Casalderrey-Solana, Jorge",
    title = "{Thermal emission of gravitational waves from weak to strong coupling}",
    eprint = "2202.05241",
    archivePrefix = "arXiv",
    primaryClass = "hep-th",
    doi = "10.1007/JHEP10(2022)049",
    journal = "JHEP",
    volume = "10",
    pages = "049",
    year = "2022"
}

@article{Caron-Huot:2009ypo,
    author = "Caron-Huot, S.",
    title = "{Asymptotics of thermal spectral functions}",
    eprint = "0903.3958",
    archivePrefix = "arXiv",
    primaryClass = "hep-ph",
    doi = "10.1103/PhysRevD.79.125009",
    journal = "Phys. Rev. D",
    volume = "79",
    pages = "125009",
    year = "2009"
}

@article{Kawai:1985xq,
    author = "Kawai, H. and Lewellen, D. C. and Tye, S. H. H.",
    title = "{A Relation Between Tree Amplitudes of Closed and Open Strings}",
    reportNumber = "CLNS-85/667",
    doi = "10.1016/0550-3213(86)90362-7",
    journal = "Nucl. Phys. B",
    volume = "269",
    pages = "1--23",
    year = "1986"
}

@article{McGreevy:2009xe,
    author = "McGreevy, John",
    title = "{Holographic duality with a view toward many-body physics}",
    eprint = "0909.0518",
    archivePrefix = "arXiv",
    primaryClass = "hep-th",
    reportNumber = "MIT-CTP-4067, NSF-KITP-09-134",
    doi = "10.1155/2010/723105",
    journal = "Adv. High Energy Phys.",
    volume = "2010",
    pages = "723105",
    year = "2010"
}

@article{Manenti:2019wxs,
    author = "Manenti, Andrea",
    title = "{Thermal CFTs in momentum space}",
    eprint = "1905.01355",
    archivePrefix = "arXiv",
    primaryClass = "hep-th",
    doi = "10.1007/JHEP01(2020)009",
    journal = "JHEP",
    volume = "01",
    pages = "009",
    year = "2020"
}

@article{Bonelli:2022ten,
    author = "Bonelli, Giulio and Iossa, Cristoforo and Panea Lichtig, Daniel and Tanzini, Alessandro",
    title = "{Irregular Liouville Correlators and Connection Formulae for Heun Functions}",
    eprint = "2201.04491",
    archivePrefix = "arXiv",
    primaryClass = "hep-th",
    doi = "10.1007/s00220-022-04497-5",
    journal = "Commun. Math. Phys.",
    volume = "397",
    number = "2",
    pages = "635--727",
    year = "2023"
}

@article{Maier_2006,
   title={The 192 solutions of the Heun equation},
   volume={76},
   ISSN={1088--6842},
   url={http://dx.doi.org/10.1090/S0025-5718-06-01939-9},
   DOI={10.1090/s0025-5718-06-01939-9},
   number={258},
   journal={Mathematics of Computation},
   publisher={American Mathematical Society (AMS)},
   author={Maier, Robert S.},
   year={2006},
   month=nov, pages={811–843} }

@article{Cai:2001dz,
    author = "Cai, Rong-Gen",
    title = "{Gauss-Bonnet black holes in AdS spaces}",
    eprint = "hep-th/0109133",
    archivePrefix = "arXiv",
    doi = "10.1103/PhysRevD.65.084014",
    journal = "Phys. Rev. D",
    volume = "65",
    pages = "084014",
    year = "2002"
}

@article{Myers:2010ru,
    author = "Myers, Robert C. and Robinson, Brandon",
    title = "{Black Holes in Quasi-topological Gravity}",
    eprint = "1003.5357",
    archivePrefix = "arXiv",
    primaryClass = "gr-qc",
    doi = "10.1007/JHEP08(2010)067",
    journal = "JHEP",
    volume = "08",
    pages = "067",
    year = "2010"
}

@article{Birmingham:1998nr,
    author = "Birmingham, Danny",
    title = "{Topological black holes in Anti-de Sitter space}",
    eprint = "hep-th/9808032",
    archivePrefix = "arXiv",
    doi = "10.1088/0264-9381/16/4/009",
    journal = "Class. Quant. Grav.",
    volume = "16",
    pages = "1197--1205",
    year = "1999"
}

@article{Mann:1997iz,
    author = "Mann, Robert B.",
    editor = "Burko, Lior M. and Ori, Amos",
    title = "{Topological black holes: Outside looking in}",
    eprint = "gr-qc/9709039",
    archivePrefix = "arXiv",
    reportNumber = "WATPHYS-TH-97-12",
    journal = "Annals Israel Phys. Soc.",
    volume = "13",
    pages = "311",
    year = "1997"
}

@article{Chamblin:1999hg,
    author = "Chamblin, Andrew and Emparan, Roberto and Johnson, Clifford V. and Myers, Robert C.",
    title = "{Holography, thermodynamics and fluctuations of charged AdS black holes}",
    eprint = "hep-th/9904197",
    archivePrefix = "arXiv",
    reportNumber = "DAMTP-1999-54, EHU-FT-9907, DTP-99-25, UK-99-5, MCGILL-99-15",
    doi = "10.1103/PhysRevD.60.104026",
    journal = "Phys. Rev. D",
    volume = "60",
    pages = "104026",
    year = "1999"
}

@article{David:2022nfn,
    author = "David, Justin R. and Kumar, Srijan",
    title = "{Thermal one point functions, large d and interior geometry of black holes}",
    eprint = "2212.07758",
    archivePrefix = "arXiv",
    primaryClass = "hep-th",
    doi = "10.1007/JHEP03(2023)256",
    journal = "JHEP",
    volume = "03",
    pages = "256",
    year = "2023"
}

@article{Emparan:1999gf,
    author = "Emparan, Roberto",
    title = "{AdS / CFT duals of topological black holes and the entropy of zero energy states}",
    eprint = "hep-th/9906040",
    archivePrefix = "arXiv",
    reportNumber = "DTP-99-37, EHU-FT-9909",
    doi = "10.1088/1126-6708/1999/06/036",
    journal = "JHEP",
    volume = "06",
    pages = "036",
    year = "1999"
}

@article{Buchel:2009sk,
    author = "Buchel, Alex and Escobedo, Jorge and Myers, Robert C. and Paulos, Miguel F. and Sinha, Aninda and Smolkin, Michael",
    title = "{Holographic GB gravity in arbitrary dimensions}",
    eprint = "0911.4257",
    archivePrefix = "arXiv",
    primaryClass = "hep-th",
    reportNumber = "UWO-TH-09-16",
    doi = "10.1007/JHEP03(2010)111",
    journal = "JHEP",
    volume = "03",
    pages = "111",
    year = "2010"
}

@article{Starinets:2002br,
    author = "Starinets, Andrei O.",
    title = "{Quasinormal modes of near extremal black branes}",
    eprint = "hep-th/0207133",
    archivePrefix = "arXiv",
    reportNumber = "INT-PUB-02-46",
    doi = "10.1103/PhysRevD.66.124013",
    journal = "Phys. Rev. D",
    volume = "66",
    pages = "124013",
    year = "2002"
}

@article{Emparan:2020inr,
    author = "Emparan, Roberto and Herzog, Christopher P.",
    title = "{Large D limit of Einstein{\textquoteright}s equations}",
    eprint = "2003.11394",
    archivePrefix = "arXiv",
    primaryClass = "hep-th",
    doi = "10.1103/RevModPhys.92.045005",
    journal = "Rev. Mod. Phys.",
    volume = "92",
    number = "4",
    pages = "045005",
    year = "2020"
}

@article{Emparan:2013moa,
    author = "Emparan, Roberto and Suzuki, Ryotaku and Tanabe, Kentaro",
    title = "{The large D limit of General Relativity}",
    eprint = "1302.6382",
    archivePrefix = "arXiv",
    primaryClass = "hep-th",
    doi = "10.1007/JHEP06(2013)009",
    journal = "JHEP",
    volume = "06",
    pages = "009",
    year = "2013"
}

@article{Bhattacharyya:2015fdk,
    author = "Bhattacharyya, Sayantani and Mandlik, Mangesh and Minwalla, Shiraz and Thakur, Somyadip",
    title = "{A Charged Membrane Paradigm at Large D}",
    eprint = "1511.03432",
    archivePrefix = "arXiv",
    primaryClass = "hep-th",
    doi = "10.1007/JHEP04(2016)128",
    journal = "JHEP",
    volume = "04",
    pages = "128",
    year = "2016"
}

@article{Soda:1993xc,
    author = "Soda, J.",
    title = "{Hierarchical dimensional reduction and gluing geometries}",
    doi = "10.1143/PTP.89.1303",
    journal = "Prog. Theor. Phys.",
    volume = "89",
    pages = "1303--1310",
    year = "1993"
}

@article{Emparan:2013xia,
    author = "Emparan, Roberto and Grumiller, Daniel and Tanabe, Kentaro",
    title = "{Large-D gravity and low-D strings}",
    eprint = "1303.1995",
    archivePrefix = "arXiv",
    primaryClass = "hep-th",
    reportNumber = "TUW-13-04",
    doi = "10.1103/PhysRevLett.110.251102",
    journal = "Phys. Rev. Lett.",
    volume = "110",
    number = "25",
    pages = "251102",
    year = "2013"
}

@article{Bzowski:2019kwd,
    author = "Bzowski, Adam and McFadden, Paul and Skenderis, Kostas",
    title = "{Conformal $n$-point functions in momentum space}",
    eprint = "1910.10162",
    archivePrefix = "arXiv",
    primaryClass = "hep-th",
    doi = "10.1103/PhysRevLett.124.131602",
    journal = "Phys. Rev. Lett.",
    volume = "124",
    number = "13",
    pages = "131602",
    year = "2020"
}

@article{Bzowski:2014qja,
    author = "Bzowski, Adam and Skenderis, Kostas",
    title = "{Comments on scale and conformal invariance}",
    eprint = "1402.3208",
    archivePrefix = "arXiv",
    primaryClass = "hep-th",
    doi = "10.1007/JHEP08(2014)027",
    journal = "JHEP",
    volume = "08",
    pages = "027",
    year = "2014"
}

@article{Armas:2020akg,
    author = "Armas, Jay and Have, Emil and Nicosia, Gianbattista-Piero",
    title = "{Null fluid/gravity correspondence}",
    eprint = "2602.20268",
    archivePrefix = "arXiv",
    primaryClass = "hep-th",
    doi = "10.1007/JHEP08(2026)220",
    journal = "JHEP",
    volume = "26",
    pages = "220",
    year = "2020"
}

@article{Armas:2025uyv,
    author = "Armas, Jay and Have, Emil and Nicosia, Gianbattista-Piero",
    title = "{Null matter and the ultrarelativistic origin of hydrodynamics at zero temperature}",
    eprint = "2509.25320",
    archivePrefix = "arXiv",
    primaryClass = "hep-th",
    month = "9",
    year = "2025"
}

\end{document}